\documentclass[10pt,journal,compsoc]{IEEEtran}
\usepackage[T1]{fontenc}
\ifCLASSINFOpdf

\else

\fi

\usepackage{amsmath}
\usepackage[cmintegrals]{newtxmath}

\usepackage[square, numbers, comma, sort & compress]{natbib}
\usepackage{fonttable}
\usepackage{url}
\usepackage{multicol}
\usepackage{booktabs}
\usepackage{lipsum}
\usepackage{colortbl}
\usepackage{multicol}
\usepackage{multirow}
\usepackage{kantlipsum} 
\usepackage{mwe} 
\usepackage{subcaption}
\usepackage{float}
\definecolor{mycolor}{rgb}{0.92,0.95,0.86}
\definecolor{mycolor1}{RGB}{255, 204, 150}
\definecolor{mycolor2}{RGB}{97, 48, 06}

\usepackage[ruled,linesnumbered]{algorithm2e}

\begin{document}
%
\title{A Survey on Machine Learning-based Misbehavior Detection Systems for 5G and Beyond Vehicular Networks}
%
%
%


\author{\IEEEauthorblockN{Abdelwahab Boualouache, Member, IEEE and Thomas Engel, Member, IEEE}
	\IEEEauthorblockA{ 
    \\ FSTM, Faculty of Science, Technology and Medicine, University of Luxembourg, Luxembourg\\ 
       Email: \{abdelwahab.boualouache, thomas.engel\}@uni.lu}\\
     }


%
%

\markboth{}%
{Shell \MakeLowercase{\textit{et al.}}: Bare Demo of IEEEtran.cls for IEEE Communications Society Journals}
%



\maketitle

\begin{abstract}

Significant progress has been made towards deploying Vehicle-to-Everything (V2X) technology. Integrating V2X with 5G has enabled ultra-low latency and high-reliability V2X communications. However,  while communication performance has enhanced, security and privacy issues have increased. Attacks have become more aggressive, and attackers have become more strategic. Public Key Infrastructure proposed by standardization bodies cannot solely defend against these attacks. Thus, in complementary of that, sophisticated systems should be designed to detect such attacks and attackers. Machine Learning (ML) has recently emerged as a key enabler to secure our future roads. Many V2X Misbehavior Detection Systems (MDSs) have adopted this paradigm. Yet, analyzing these systems is a research gap, and developing effective ML-based MDSs is still an open issue. To this end, this paper present a comprehensive survey and classification of ML-based MDSs. We analyze and discuss them from both security and ML perspectives. Then, we give some learned lessons and recommendations helping in developing, validating, and deploying ML-based MDSs.  Finally, we highlight open research and standardization issues with some future directions.

\end{abstract}


%
\IEEEpeerreviewmaketitle


\section{Introduction} \label{intro}

The emergence of the fifth-generation mobile communications networks (5G) has brought a technological revolution to the world, as it provides ultra-low latency, ultra-reliability, high bandwidth, and scalable coverage \cite{yin2020multiplexing}.  As a part of the vision of 5G, V2X communications are witnessing tremendous advances. 5G-V2X aims to ensure road safety, avoid traffic congestion, and provide a better driving experience for users during their journey \cite{hussain2018autonomous}.  However, facing a huge vector of attacks, 5G-V2X is suffering from security and privacy issues, which can lead to hazardous situations for drivers and passengers. These issues have been taken special attention by research communities since early research investigations on V2X \cite{lu20195g}.
Extensive research works have been carried out to protect V2X. Several cryptography solutions have been proposed for thwarting V2X attacks \cite{wang2020physical, azam2021comprehensive,yang2021blockchain}. In addition, standardization bodies have designed a Public Key Infrastructure (PKI) to offer V2X security services, especially authentication, integrity, and confidentiality \cite{etsi8}. Standard specifications define not only messages formats but also all cryptography tools to sign and encrypt V2X messages. However, although an important vector of attacks has been avoided using these solutions, more aggressive attacks still persist.   More specifically, internal attacks such as denial of service, position, and message droppings, falsification pose a real danger since attackers are already authenticated members, which makes them resistant to cryptographic solutions \cite{gyawali2021deep}. In addition, attackers have become more intelligent and strategic to overcome the defense lines \cite{9612604}. In this context, Misbehavior Detection Systems (MDSs) have been proposed as complementary to PKI to detect such attacks, and then exclude attackers from the 5G-V2X system.  However, detecting these attackers is challenging and require employing sophisticated and intelligent detection mechanisms. 

Machine learning (ML) has recently emerged a key enabler of intelligence for our future networks. It becomes obvious that ML algorithms will be of the pillar of 5G and beyond and 6G mobile networks \cite{mao2021ai}. In addition,  ML algorithms have already proven success in the areas of network security \cite{buczak2015survey}.  Consequently,  several ML-based MDSs have been proposed to detect attacks 
on 5G-V2X.

\begin{figure}[H]
\centering
\includegraphics[width=9cm,height=6cm]{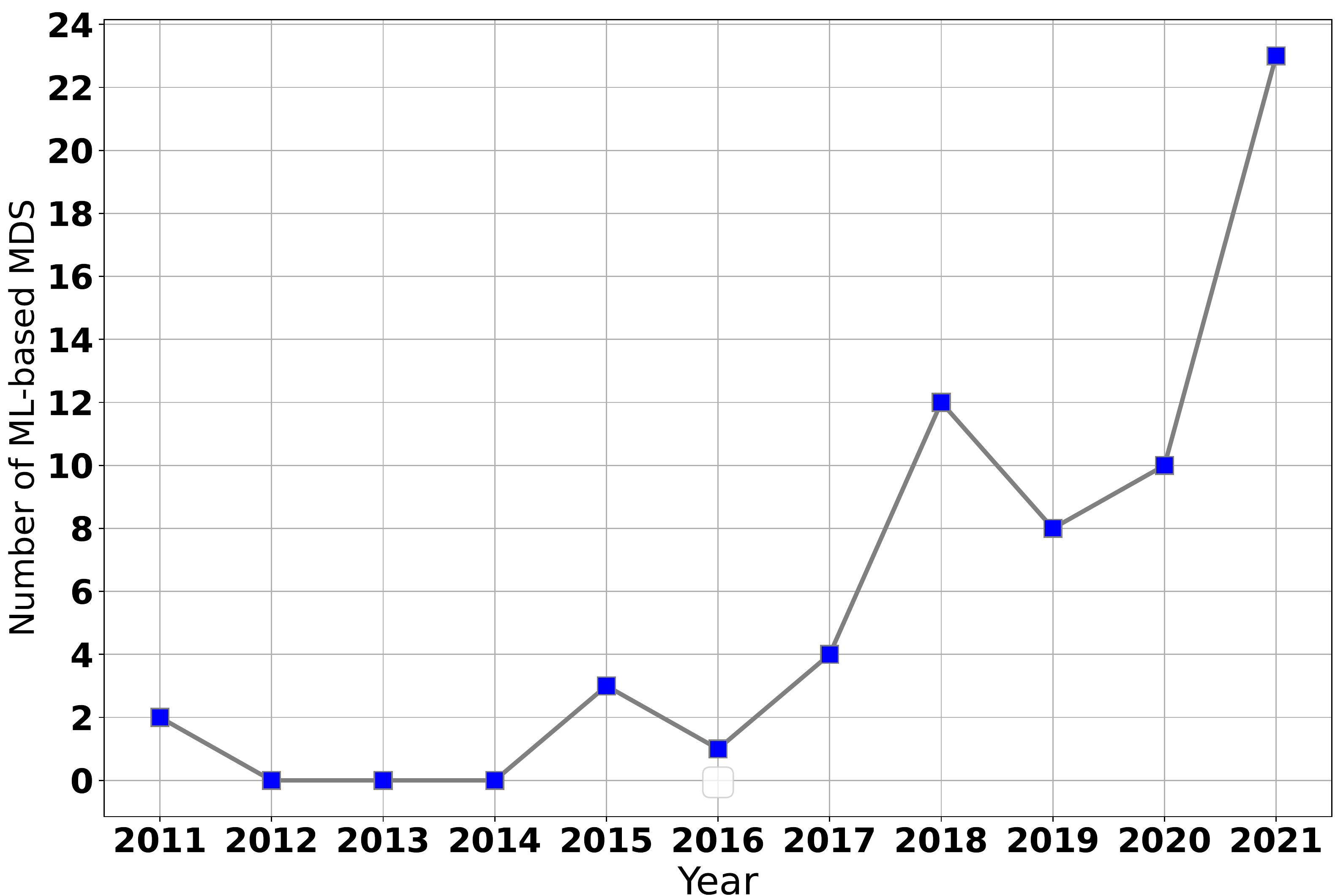}
\caption{Number of published ML-based MDSs per year}
\label{fig:pubMDSperYear}
\end{figure}

Figure~\ref{fig:pubMDSperYear} is a result of our quantitative study presented in Table~\ref{tab:MDSvsAtk}. It shows the number of published ML-based MDSs per year. As we can see, recent years have witnessed a notable increase in the number of proposed ML-based MDSs. This is due to the trust of the research communities in ML for providing efficient and evolutive MDSs \cite{tang2021comprehensive}. However, analyzing ML-based MDSs is still a research gap. The survey comes to fill this gap and to complement ongoing research and standardization activities on MDS \cite{etsi3}. This survey analyzes existing ML-based MDSs not only from security but also from ML perspectives. Thus, establishing analysis guidelines and presenting learned lessons and recommendations for future ML-based MDSs. It also identifies research and standardization open gaps, to which focus and priority should go to succeed ML-based MDSs deployment.

	\subsection*{\textbf{Relevant surveys}}
	
	Several surveys have been conducted on security and privacy in 5G vehicular networks. The authors of~\cite{sun2021survey} highlighted security challenges in the 5G-V2X environment. They also identified various cyber-security risks and vulnerabilities and analyzed corresponding defense strategies for securing connected vehicles. The authors of~\cite{dibaei2020attacks} classified the available defenses mechanisms into four categories: cryptography, network security, software vulnerability detection, and malware detection. The authors of \cite{sakiz2017survey} surveyed possible attacks and the corresponding  detection mechanisms. The authors of \cite{arshad2018survey,almalki2020review} reviewed detection schemes of data falsification attacks. The authors of \cite{van2018survey} gave a clear definition of V2X misbehavior. They also reviewed different MDS and provided a comprehensive classification of the existing MDS. However, all previous surveys have reviewed MDS in general without focusing on ML aspects. The authors of  \cite{dibaei2021investigating}  discussed the role of ML in enabling efficient cybersecurity defense mechanisms in 5G-V2X. The authors of \cite{talpur2021machine} classified the ML techniques according to their use in V2X applications and discussed approaches and working principles of these ML techniques in addressing various security challenges. The authors of \cite{rajbahadur2018survey} only focused on the MDS that use a subset of ML techniques, unsupervised anomaly detection techniques. Moreover, the MDS of only three communication attacks are surveyed: false information, black grey and wormhole attacks, and DoS. The authors of \cite{alrehan2019machine} surveyed ML-based MDS detecting only DDoS attacks. Finally, the authors of \cite{gonccalves2019systematic} presented a Systematic Literature Review (SLR) for several ML-based MDS for 5G-V2X along with their ML algorithms, architectures, and datasets. However, although some ML-based MDSs have been covered in the previous surveys~\cite{talpur2021machine, rajbahadur2018survey, alrehan2019machine, gonccalves2019systematic}, the coverage of the existing ML-based MDSs is still limited and lacking from deep analysis. Thus, to complement these efforts and in contrast to previous surveys, this paper particularly presents a comprehensive survey of existing ML-based MDSs for 5G-V2X. The paper deeply analyzes these solutions to identify their strengths and weaknesses along with research and standardization gaps. To the best of our knowledge, we are the first to propose such a survey. We hope that this survey will build guidelines to select best ML-based MDSs to implement in the near deployment of 5G-V2X and shape future research directions in this topic.


	\subsection*{\textbf{Contributions}}
	
	The main contributions of this paper can be summarized as follows:
	\begin{itemize}
		\item We survey and elaborate taxonomy of machine learning-based misbehavior detection systems.
		\item We analyze and discuss the presented solutions.
		\item We present lessons learned and recommendations for developing, evaluating, and deploying ML-based MDSs.
		\item We highlight open research and standardization issues on the topics.
	\end{itemize}
	
\begin{figure*}[!ht]
		\centering
		\includegraphics[width=18cm,height=8cm]{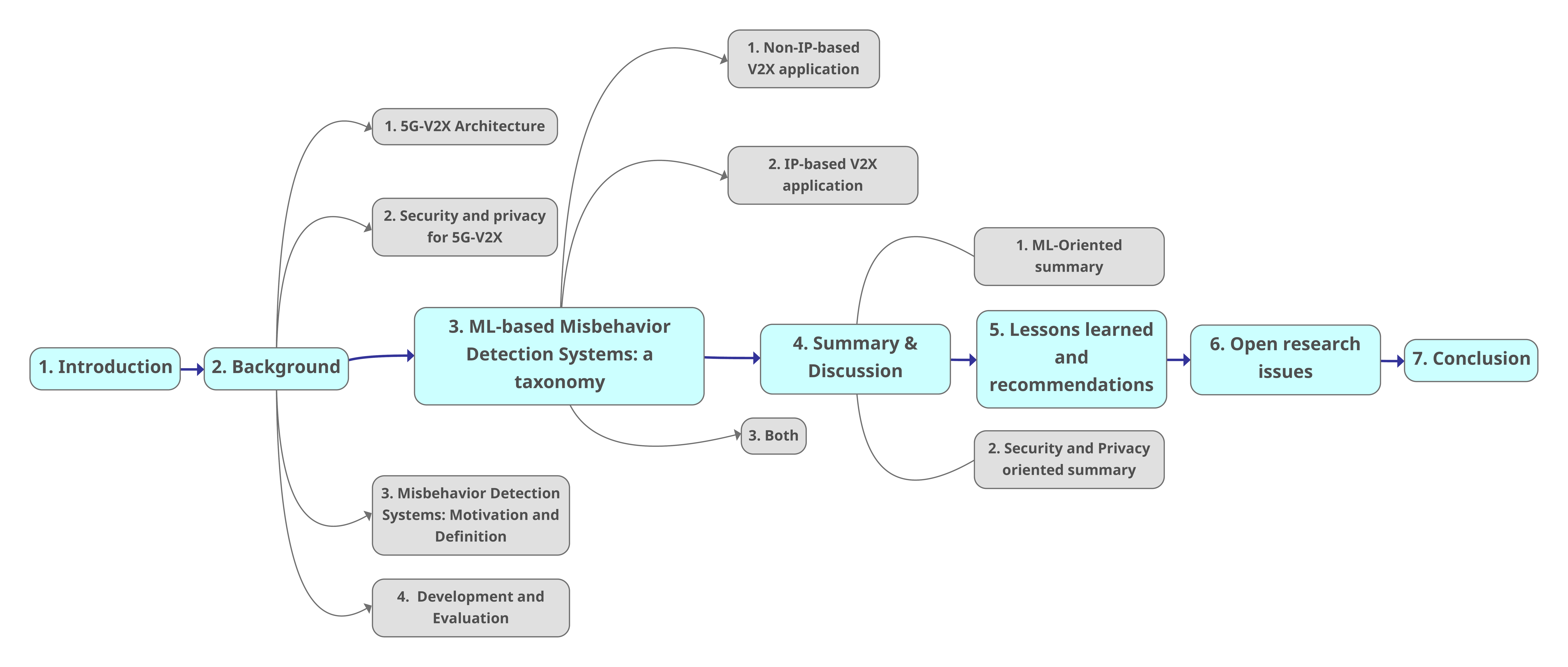}
	    \caption{Survey's roadmap}
		\label{fig:roadmap}
\end{figure*}

	The rest of the paper is organized as follows. In Section~\ref{sec:background}, we present some necessary background information. A taxonomy of machine learning-based misbehavior detection systems is presented in Section~\ref{sec:taxonomy}. In Section~\ref{sec:summary}, we analyze and discuss the presented ML-based MDSs. Lessons learned and recommendations are discussed in Section~\ref{sec:lessons}. Open research issues are given in Section~\ref{sec:issues}. Finally, Section~\ref{sec:conclusion} concludes this survey.
The roadmap of this survey is given in Figure~\ref{fig:roadmap}.

\section{Background } \label{sec:background}

The purpose of this section is to give the reader the necessary background information to understand the research presented in this paper. This section is divided into four subsections. Firstly, we describe the architecture of 5G-V2X. Then, we overview security requirements, attacker models, and attacks on 5G-V2X. After that, we motivate and define the MDS. Finally, we describe the development and evaluation elements of ML-based MDSs. Abbreviations used throughout the paper are described in Table~\ref{fig:abrv}.

\begin{table}[!h]
	\caption{Abbreviations used throughout the paper.}
	\label{fig:abrv}
	\begin{normalsize}
		\begin{tabular}{|p{1.4cm}|p{6.6cm}|}
			\hline \textbf{Abbr}  & \textbf{Description} \\
			\hline 3GPP & The 3rd Generation Partnership Project \\
			\hline 5G &  The 5th generation mobile network \\
			\hline AUC &  Area Under the Curve \\
			\hline AoA &  Angle of Arrival \\
			\hline CNN & Convolutional Neural Network \\
			\hline DR &  Detection Rate \\
			\hline DS &  Decision Stump \\
			\hline DDoS &  Distributed Denial of Service\\
			\hline DoS & Denial of Service \\
			\hline ET & Extra Tree \\
			\hline ETSI & European Telecommunications Standards Institute\\
			\hline FL &   Federated learning \\
			\hline FN &  False Negative \\
			\hline FNR & False Negative Rate \\
			\hline FP & False Positive \\
			\hline FPR & False Positive Rate \\
			\hline GAN& Generative Adversarial Network \\
			\hline GPS & Global Positioning System  \\
			\hline GRU &  Gated Recurrent Unit  \\
			\hline IBL &  Instance Based Learning \\
			\hline CNN & Convolutional Neural Network \\
			\hline IP &  Internet Protocol \\
			\hline KNN & k-Nearest Neighbors \\
			\hline LGBM & Light Gradient Boosting Machine\\
			\hline LR & Logistic Regression\\
			\hline LSTM & Long Short-Term Memory  \\
			\hline LTE & Long-Term Evolution  \\
			\hline MDS & Misbehavior Detection Systems \\
			\hline ML & Machine Learning  \\
			\hline NB & Naive Bayes   \\
			\hline NN & Neural Networks \\
			\hline NR & New Radio \\
			\hline PKI & Public Key Infrastructure \\
			\hline RF & Random Forest  \\
			\hline RNN & Recurrent Neural Network  \\
			\hline ROC & Receiver Operator Characteristic \\
			\hline RSSI & Received Signal Strength and interference    \\
			\hline RSU & Roadside Unit   \\
			\hline SDN & Sotware Defined Networking  \\
			\hline SST & Singular Spectrum Transformation   \\
			\hline SVM & Support-vector machine \\
			\hline TCP & Transmission Control Protocol   \\
			\hline TN & True Negative \\
			\hline TNR & True Negative Rate \\
			\hline TP & True Positive  \\
			\hline TPR & True Positive Rate  \\
			\hline UDP & User Datagram Protocol  \\
			\hline V2I & Vehicle-to-Infrastructure \\
			\hline V2N & Vehicle-to-Network \\
			\hline V2P & Vehicle-to-Pedestrian \\
			\hline V2V & Vehicle-to-Vehicle\\
			\hline V2X & Vehicle-to-Everything  \\
			\hline VUE & Vehicular User Equipment \\
			\hline
		\end{tabular}
	
	\end{normalsize}

\end{table}

\subsection{5G-V2X Architecture}
This section describes the main building blocks of 5G-V2X.

\subsubsection{Architecture}
V2X communications aim to provide a safer and comfortable driving experience.  Two technologies have been developed to enable V2X communications. These technologies, currently seen as alternatives, are IEEE 802.11p (ITS-G5 in Europe) and Cellular vehicle-to-everything (C-V2X). However, ITS-G5 has known a slow development on a wide scale in the favour of C-V2X, which is witnessing a significant growth led by 3GPP \cite{campolo20215g}. C-V2X technology is already part of the completed 3GPP Long-Term Evolution (LTE) Releases 14 and 15~\cite{3gpp2}. It is designed to support ultra-lately and ultra-reliable V2X use case groups specified Release 16~\cite{3gpp1}. V2X communication types are classified as follows \cite{3gpp3}: (i) vehicle-to-vehicle (V2V) for direct communications between Vehicular user equipment (VUEs); (ii) vehicle-to-infrastructure (V2I) for communications between vehicles and the RSUs, which can be deployed as gNodeBs or in a standalone devices; (iii) vehicle-to-pedestrian (V2P) between VUEs and Vulnerable Road Users (VRUs) such as  pedestrians and bikers; and (iv) vehicle-to-network (V2N) for communications with remote servers and cloud-based services reachable through the cellular infrastructure. The enhancement of 3GPP to support C-V2X communications concerns both the radio access network (the New Radio) and the core network.

\begin{itemize}
    \item \textbf{5G NR V2X}: Several enhancements are introduced in Release 16 within New radio (NR) to support V2X applications’ demands in terms of latency and reliability~\cite{3gpp1}. These enhancements are ranging from introducing more disruptive radio technologies (e.g., flexible waveforms) to improving modes 3 and 4 in specific communication modes (e.g., multicast and groupcast). V2X NR covers both the PC5 and LTE-Uu radio interfaces. The PC5 radio interface (sidelink) is used for V2V, V2P, and V2I communications, bypassing the cellular infrastructure. In the absence of a cellular network, the 5.9 GHz band is employed to ensure ultra-high availability in all regions, regardless of the mobile network operator (MNO). Two direct communication modes are supported by PC5 Mode 3 (scheduled), and Mode 4 (autonomous). Mode 3 only operates in-coverage of a gNodeB, which is in charge of the allocation of radio resources. Instead, Mode 4 can operate both in- and out-of-coverage of an eNodeB where the allocation of radio resources is agreed between vehicles without support for the infrastructure. V2N communications occur over the conventional cellular Uu interface operating in the licensed spectrum.  This interface has been modified to handle both unicast and multicast V2X communications with fewer changes that enable efficient V2X information sharing to meet the latency requirements of V2X applications. 
    \item \textbf{5G Core Network}:
   The 5G core network is designed to enable mobile data connectivity and support various verticals leveraging emerging technologies, such as Software-Defined Networking (SDN) and Network Functions Virtualization (NFV). By separating the user plane function (UPF) from the control plane function (CPF) the 5G core becomes scalable and flexible. The building blocks of the 5G core are a set of virtual network functions, including authentication server function (AUSF), access and mobility management function (AMF), user plan function (UPF), session management function (SMF), network slice selection function (NSSF), uniﬁed data management (UDM), application function (AF), network repository function (NRF), network exposure function (NEF), and security edge protection proxy (SEPP). 
\end{itemize}

\begin{figure}[!ht]
	\centering
	\includegraphics[width=0.5\textwidth]{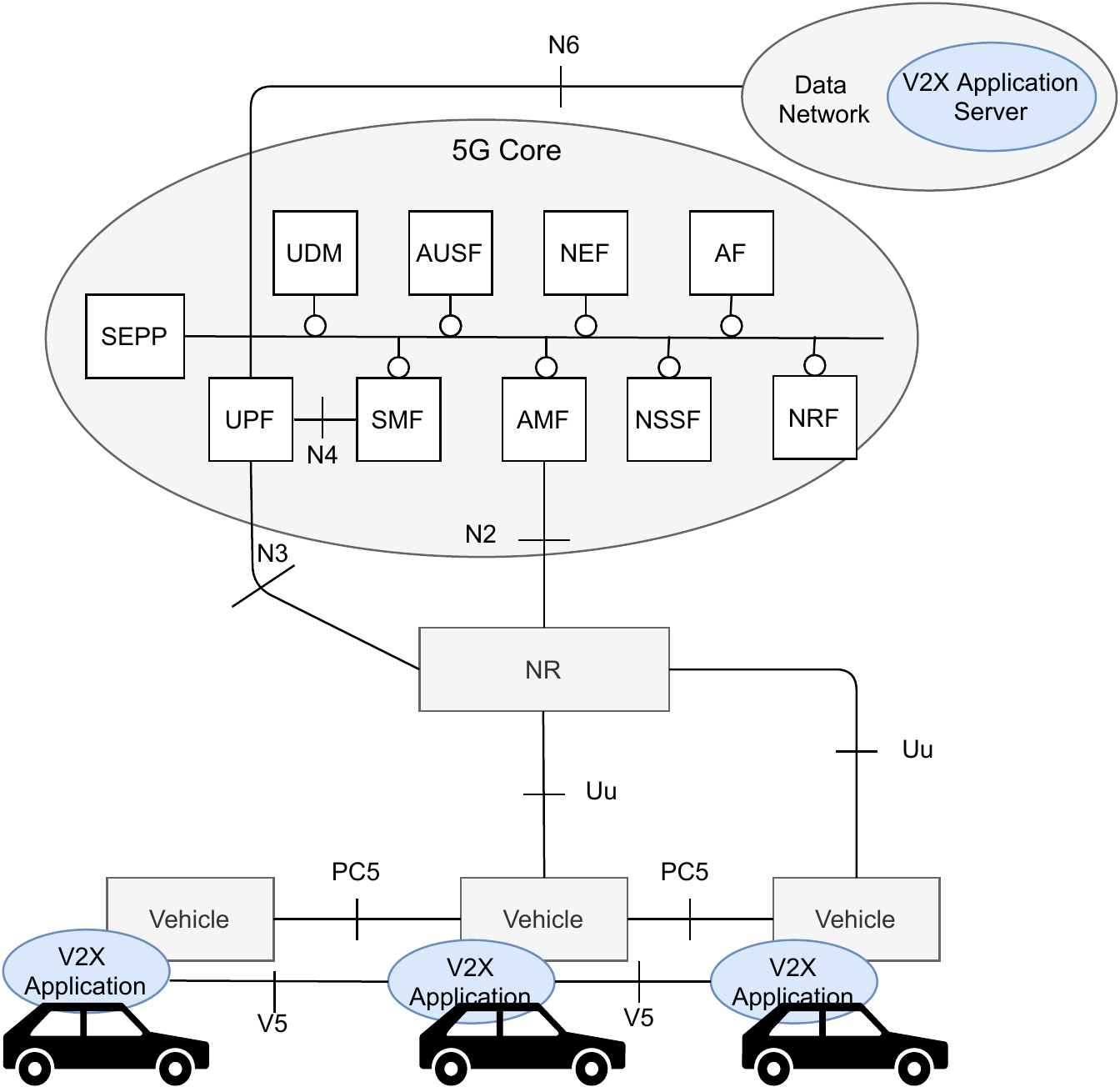}
 	\caption{5G-V2X Architecture}
	\label{fig:v2x_architecture}
\end{figure}

Figure~\ref{fig:v2x_architecture} shows a high-level view of the 5G-V2X architecture for V2X communication over PC5 and Uu reference points.

\subsubsection{Use case groups}

3GPP technical report (TR) 22.886~\cite{3gpp4} presents a comprehensive description of the envisaged 5G-V2X use case groups, which are given as follows~\cite{9345798}:

\begin{itemize}
    \item \textbf{Vehicles Platooning}: this group includes use cases that enable forming groups of vehicles in platoons while maintaining their functioning through the periodic exchange of messages. 
    
    \item \textbf{Advanced Driving}: This group comprises use cases that allow for semi- or fully automated driving while ensuring traffic efficiency and road safety. 
    
    \item \textbf{Extended Sensors}: This group is to improve the perception of vehicles through the exchange of data collected from different data sources such as local sensors, RSUs, and VRUs.
    
    \item \textbf{Remote Driving}: This group enables to drive connected vehicles remotely. It includes, for example, remote assistance of beginner drivers to overcome difficult road situations or automated vehicles to perform complex maneuvers.

\end{itemize}

\subsubsection{Technologies and Standards}
Two technologies have been developed to enable direct information sharing between vehicles. These technologies, currently seen as alternatives, are ETSI ITS-G5, as commonly referred to in Europe and C-V2X~\cite{bazzi2020co}.  This subsection describes the protocol stacks of ETSI ITS-G5 and C-V2X respectively. 

\subsubsection*{ETSI ITS-G5 protocol stack}
ETSI ITS-G5 is based on a physical layer and MAC layer defined in the IEEE 802.11p protocol. The 802.11p modifies the physical and MAC layers of 802.11a to be adapted for V2X communications in a frequency band from 5.85 to 5.925 GHz, which is segmented into seven channels of 10MH each. ITS-G5  standard uses the Decentralized Congestion Control (DCC) protocol to minimize the probability of radio channel congestion. As shown in Figure~\ref{fig:protocol_stack_cv2x} (a), for the IP-based applications (non-safety applications, ITS-G5 uses the IP for the network layer and the UDP/TCP for the transport layer. On the other hand, for non-IP-based applications (safety applications), ITS-G5 uses the Geonetworking protocol to enable packets' routing based on the geographic position of vehicles in the network layer \cite{etsi1}, while the Basic Transport Protocol (BTP) is used to offer point-to-point connectionless network transport service in the transport layer \cite{etsi2}. ITS-G5 also introduces the facilities layer between the transport layer and the applications layer, where several messages were defined \cite{abunei2017implementation}. For example, the CAM (Cooperative Awareness Message) was defined for the periodic messages and the DENM (Decentralized Environmental Message) was defined for the event messages.

\subsubsection*{C-V2X User Plan Stack}

\begin{figure*}[!ht]
	\centering
	\includegraphics[width=1\textwidth]{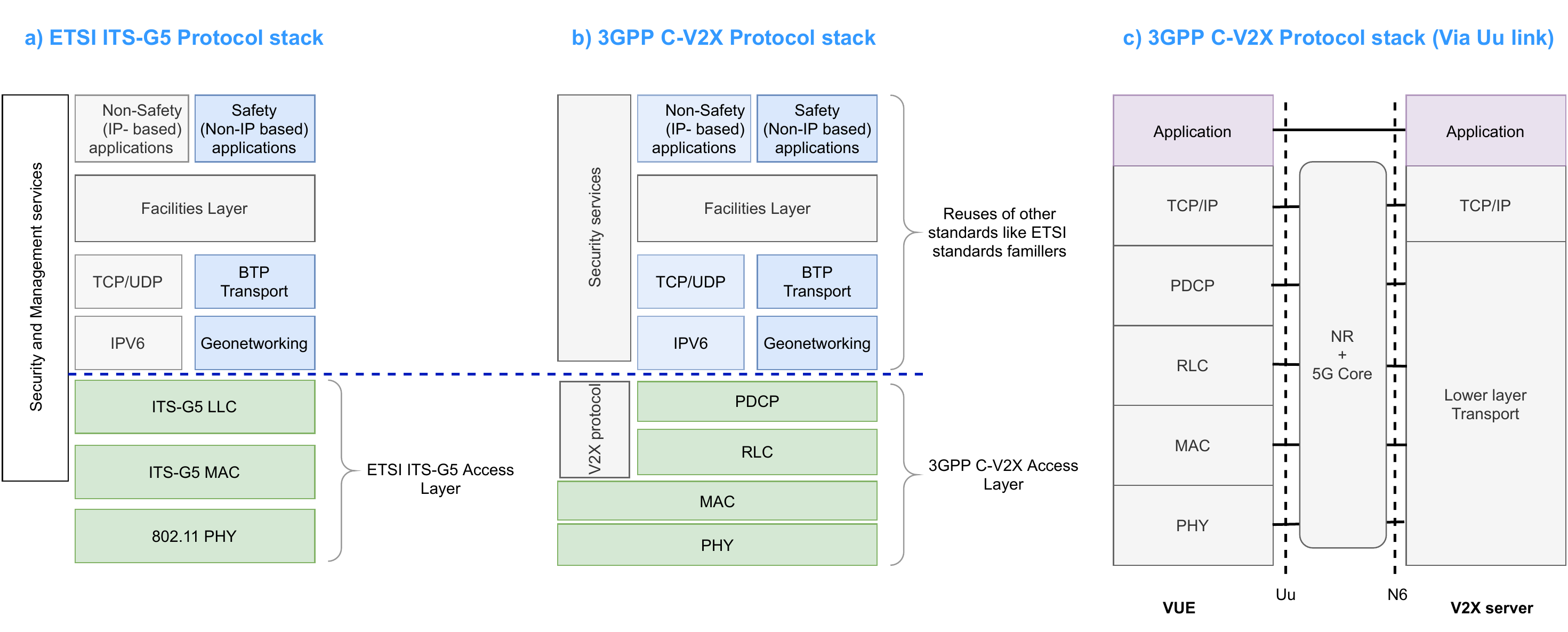}
 	\caption{V2X Protocol Stacks \cite{Qualcomm1,Qualcomm2}}
	\label{fig:protocol_stack_cv2x}
\end{figure*}

As shown in Figure~\ref{fig:protocol_stack_cv2x} (b), the protocol stack of C-V2X via PC5 interface is mainly based on the 3GPP Releases for the low layers (PHY, MAC, RLC, and PDCP) and reuses the layer stacks from IEEE and ETSI for the upper layers (network and transport layers) \cite{Qualcomm1}. The physical layer (PHY) transmits data on the sidelink, exploiting 10MHz or 20MHz bandwidths at the 5.9GHz radio frequency band.  The media access control (MAC) layer implements the blind hybrid automatic repeat request (HARQ) without feedback. The Radio Link Control (RLC) layer is in charge of delivering service data units in sequence, as well as segmenting and reassembling them.  The packet data convergence protocol (PDCP) sublayer separates 3GPP radio access protocol layers from those related to V2X applications \cite{Qualcomm2}. As shown in Figure~\ref{fig:protocol_stack_cv2x} (c) The protocol stack of C-V2X via Uu link for V2N is common for communications in 5G architecture.

\subsection{Security and privacy for 5G-V2X}
This section is divided into three subsections: security requirements, attacker models, and attack classification. 

\subsubsection{Security requirements} \label{sec:securityservices}
5G-V2X communications are subject to a large range of attacks and cyber-threats, which can have major negative consequences for the integrity and functionality of the 5G-V2X system, potentially putting drivers' lives at risk. This subsection discusses the security services for 5G-V2X networks. 

\begin{itemize}
    \item \textbf{Authentication}: to prevent unauthorized users from injecting false messages across the 5G-V2X network, entity authentication is required. Apart from entity authentication, data authentication is also important to verify that the received data is not tampered with or replayed.
    
    \item \textbf{Integrity}: Data integrity is a must to protect drivers from malicious V2X participants, which can generate rogue messages that can affect network operations.
    
    \item \textbf{Availability}: Availability ensures that V2X messages are delivered not just to all of the intended recipients, but also at the right moment. It also ensures the continuity of V2X services.
    
    \item \textbf{Confidentiality}: confidentiality ensures that only authorized V2X participants can have access to the exchanged data.
    
    \item \textbf{Non-repudiation}: non-repudiation is necessary to prevent legitimate participants from denying the transmission or the content of their messages. 
    
    \item \textbf{Access control}: access control is necessary to ensure the system's reliability and security. To protect the safety of legitimate V2X participants, misbehaving V2X nodes should quickly be revoked from the system. 
    
    \item \textbf{Privacy}: privacy protection is an important factor in public acceptance and the successful deployment of 5G-V2X.  Three classes of the privacy protection in V2X communication system can be distinguished: (i) the identifier privacy protection (ii) the location privacy protection, and (iii) the protection of the data exchanged. 
    
\end{itemize}

\subsubsection {Attacker model}

\label{sec:adversarymodel}

Because of the V2X system's intricacy, different types of adversaries can launch various attacks. The types of potential adversaries in V2X have been thoroughly examined in the literature~\cite{raya2}, and the following types of attackers have been identified: 

\begin{itemize}
\item	\textbf{Global vs. Local}: A global attacker has a wider coverage of the V2X system than a local attacker. It can then eavesdrop on every message sent out by any vehicle. 
	
\item    \textbf{Active vs. Passive}: an active attacker can alter or inject messages in the V2X system, while a passive attacker can only eavesdrop messages.
    
\item    \textbf{Internal vs. External}: an internal adversary is an authenticated participant of V2X system, while external adversary is an intruder.
    
\item    \textbf{Malicious vs. rational}: a malicious attacker aims to damage the V2X system, without caring about its interests, while a rational attacker aims to achieve its interests while performing attacks 
\end{itemize}

\subsubsection{Attack classification}

\begin{figure*}[!ht]
	\centering
	\includegraphics[width=18cm,height=8cm]{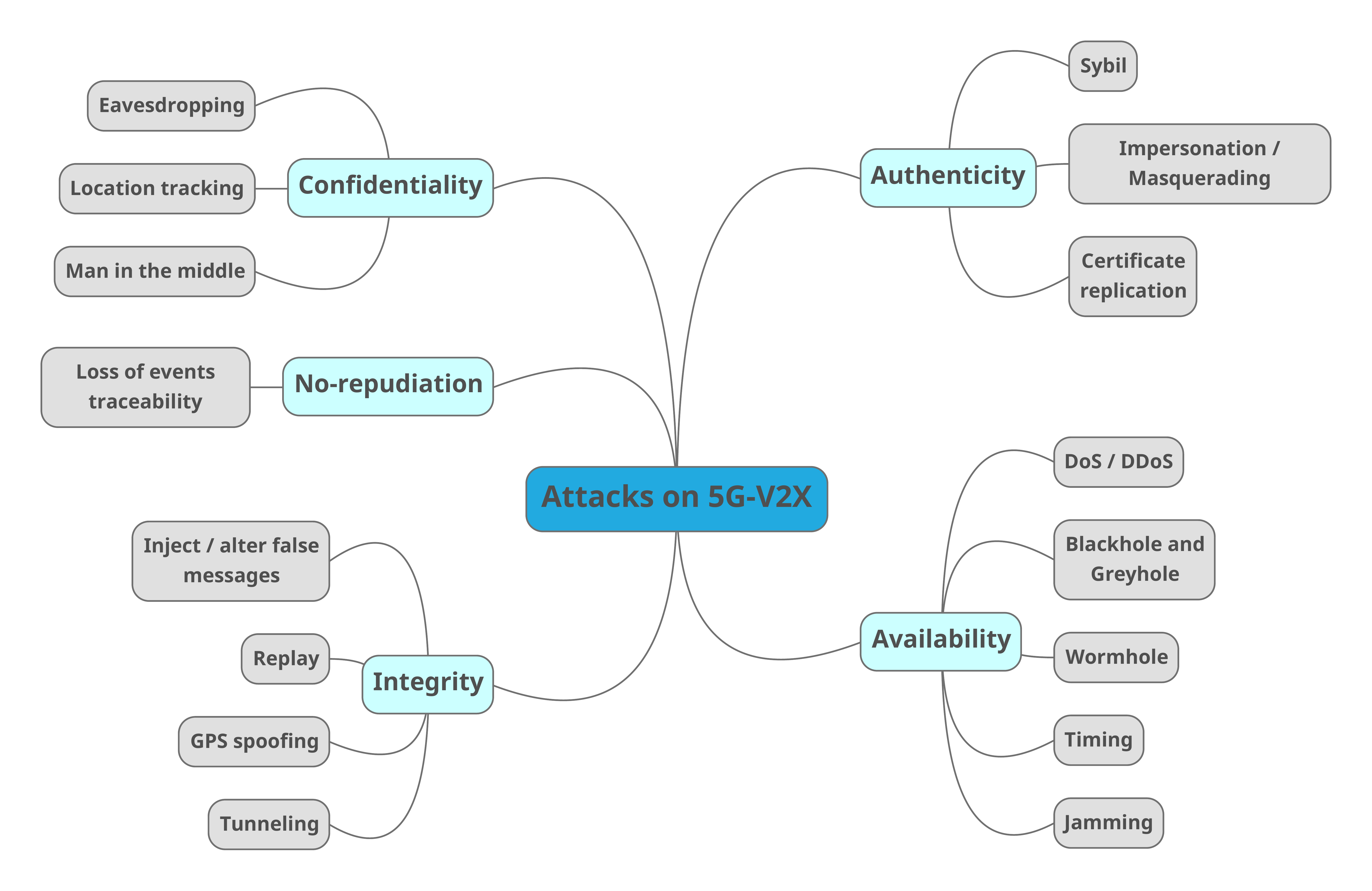}
 	\caption{Attack Classification}
	\label{fig:attack_classification}
\end{figure*}

In the following, we classify and describe attacks on 5G-V2X. Figure~\ref{fig:attack_classification} shows an overview of these attacks while Table~\ref{tab:attapp} specifies the applications targeting by these attacks and type of attacker (internal or external) that can launch them.

\subsubsection*{1) Attacks on Authenticity}
\begin{itemize}

    \item \textbf{Sybil}: this attack particularly concerns non-IP-based V2X applications (safety applications) where vehicles use multiple identifiers to protect their location privacy. However, these identifiers can also be exploited as Sybils, for example, to inject false information in V2X system for altering the perception of vehicles or to create the illusion of traffic congestion.
    
    \item \textbf{Impersonation or masquerading}: where the attacker exploits a valid identity to obtain V2X access for launching more advanced attacks and steal private information. More specifically, this attack mainly exploits the vulnerabilities in IP-based applications to get remote access to the V2X node through a multi-stage process starting to probing and port scanning to network and application layers exploits such as Malware, SQL injection, DNS poisoning.

    \item \textbf{Certificate replication}: in which malicious V2X nodes try to hide their identities by utilizing replicated certificates. Once a certificate has been blacklisted, it will no longer be used and will be removed.
    

\end{itemize}

\subsubsection*{2) Attacks on Integrity}

\begin{itemize}
    \item \textbf{Inject/Alter false messages}: In this attack, malicious nodes send wrong information (e.g. position, speed, ..etc) to honest vehicles, which may put them in dangerous situations. This attack could be more likely in non-IP-based applications.
    
    \item \textbf{Replay}: In this attack, malicious V2X nodes replay messages captured at different times and show them as generated by original senders. 
    
    \item \textbf{GPS spoofing}: malicious nodes deceive GPS receivers of other V2X nodes by re-transmitting real GPS signals captured elsewhere at a different time or by transmitting inaccurate GPS signals. 
    
    
    \item \textbf{Tunneling}: In this attack, the attacker controls at least two V2X nodes to establish a tunnel between them, and hence, it can inject false data from one place to another. Tunneling can be seen as a special case of a false messages injection attack.

    
    \item 
    
\end{itemize}

\subsubsection*{3) Attacks on Availability}

\begin{itemize}
    \item \textbf{Deny of service (DoS)/Distributed Deny of Service (DDoS):} where the attacker prevents vehicles from having normal access to network services.  DDoS attack (Distributed Denial of Service) is a variant of DoS attack that involves a set of malicious V2X nodes.  Both IP-based and non-IP-based  V2X applications are vulnerable to DoS attacks. In non-IP-based applications, the DoS attacks can be achieved by increasing the frequency of the periodic messages, whereas in IP-based applications the DoS attack can be performed in different levels such as UDP flooding and ARP flooding.  
    
    \item \textbf{Blackhole and Greyhole}: In these attacks, malicious V2X nodes stop disseminating received messages to the neighboring V2X entities. While Greyhole only selected messages are dropped, Blackhole attacker drops all received messages. These attackers particularly concern both IP-based routing protocols (eg. AODV) and position-based routing protocols (e.g. GeoNetworking).

    \item \textbf{Wormhole}: similar to the tunneling attack, in the wormhole attack, attackers establish a tunnel between malicious V2X nodes to conducting a DOS attack disrupting IP-based routing protocols. 

    \item \textbf{Timing attack}: In this attack, malicious V2X nodes intentionally delay forwarding the received messages to the next nodes in dissemination and routing protocols. This attack is very dangerous especially in time-sensitive safety-related applications
    
    \item \textbf{Jamming attack}: where the attacker generates signals to corrupt the data or jam the radio channel. Both ETSI ITS-G5 and C-V2X standards are vulnerable to this attack. 

\end{itemize}

\subsubsection*{4) Attacks on Confidentiality}
\begin{itemize}
    \item \textbf{Eavesdropping}: where the attacker gathers data from the V2X network to extract information from which it can benefit.  

\item \textbf{Location tracking}: where the attacker exploits unencrypted safety-related messages, include mobile information to track the trajectories of their victims.

\item \textbf{Man in the middle attack}: in which the attacker establishes separate connections with the victims and passes messages between them to give the impression that they are in direct communication, but in fact, all conversations between the two victims are intercepted. 

\end{itemize}

\subsubsection*{5) Attacks on Non-repudiation}

\begin{itemize}
    \item \textbf{Loss of events traceability}: in which the attacker performs a series of actions to help in the denial of specified events. These actions mostly involve deleting its traces or causing confusion for the auditing entity. 
\end{itemize}



\begin{table*}[!ht]
\caption{Attacks vs. (V2X Applications and Attacker Type) }
\label{tab:attapp}
\resizebox{18cm}{!}{
\begin{tabular}{|c|c|c|c|c|c|}
\hline
Security Service                 & Attack                        & Non-IP-based V2X application                                                      & IP-based V2X   applications                                                                          & External & Internal \\ \hline
\multirow{3}{*}{Authenticity}    & Impersonation or masquerading & X                                                                               & X                           & X        & X     \\ \cline{2-6} 
                                 & Sybil attack                  & X                                                                                 &                                                                                                      &          & X        \\ \cline{2-6} 
                                 & Certificate replication       & X                                                                                 & X                                                                                                    &          & X        \\ \hline
\multirow{4}{*}{Integrity}       & Inject/ Alter false messages  & X  & X                                                                                                  &          & X        \\ \cline{2-6} 
                                 & Replay                        & X                                                                                 & X                                                                                                    & X        & X        \\ \cline{2-6} 
                                 & GPS spoofing                  & X                                                                                 & X                                                                                                    & X        & X        \\ \cline{2-6} 
                                 & Tunneling                     &                                                                                   & X                                                                                                    &          & X        \\ \hline
\multirow{4}{*}{Availability}    & Deny of service (DoS)         & X            & X & X      & X        \\ \cline{2-6} 
                                 & Blackhole and Greyhole        & X                                                                               & X                                                                                                    &          & X        \\ \cline{2-6} 
                                 & Jamming attack                & X                                                                                 & X                                                                                                    & X        & X        \\ \cline{2-6} 
                                 & Wormhole                      &                                                                                   & X                                                                                                    &          & X        \\ \hline
\multirow{3}{*}{Confidentiality} & Eavesdropping                 &  X                                                                                 &     X                                                                                                 & X        & X        \\ \cline{2-6} 
                                 & Man in the middle             & X                                                                                 &                                                                                                      &          & X        \\ \cline{2-6} 
                                 & Location tracking             & X                                                                                 &                                                                                                      & X        & X        \\ \hline
Non-repudiation                  & Loss of events traceability   & X                                                                                 & X                                                                                                    &          & X        \\ \hline
\end{tabular}
}
\end{table*}

\subsection{Misbehavior detection systems: Motivation and Definition}

Several cryptography solutions have been proposed for thwarting V2X attacks. More specifically, standardization bodies have designed a Public Key Infrastructure (PKI) to offer V2X security services, especially authentication, integrity, and confidentiality \cite{etsi8}. Standard specifications define not only messages formats but also all cryptography tools to sign and encrypt V2X messages \cite{etsi9}. However, although an important vector of attacks has been avoided using these solutions, attacks are still be performed especially from internal attackers. In this context, misbehavior detection systems have been proposed as complementary to PKI to detect attacks, and then exclude attackers from the V2X system. 

In this survey,  misbehavior refers to both faulty and malicious actions. Faulty nodes are the nodes that generate incorrect data without malicious intent. For example, a malfunctioning vehicle onboard GPS sensor can provide incorrect position data due to damage or other technical issues. On the other hand, malicious nodes or attacker nodes are those nodes that transmit erroneous messages with malicious intent. Misbehavior detection systems can be classified into three groups described as follows~\cite{van2018survey}:

\begin{enumerate}
    \item \textbf{Node-centric}: check if the node's behaviors (e.g. message frequency and the ratio between the received and forwarded packets) are in line with protocol specifications. They can be divided into two classes: (i) behavior-based: in which the attacker is detected in case of abnormal actions (e.g., message dropping), and (ii) Trust-based: in which trust values are assigned to V2X nodes. An attacker is detected if its trust value drops below a certain predefined threshold

    \item \textbf{Data-centric}: focus on the plausibility and consistency of data, which can be individually or collaboratively verified by nodes. These systems can also be divided into two classes:  (i) Plausibility based: which use plausibility checks to decide on the correctness of data such as the received speed and position, and (ii) Consistency based: which inspect the relations between message to decide on the trustworthiness of newly received messages. For example, checking the difference between two received positions given a constant speed.
    
    \item \textbf{Hybrid}: adopt a combined approach that uses a node-centric system to evaluate nodes according to the correctness of the exchanged data, while the correctness of data is veriﬁed using a data-centric mechanism.
     
\end{enumerate}

ML can be used in all these categories of MDS. Indeed, both statistics about the V2X (node-centric) and content of exchanging messages (data-centric) can be serving to data to feed ML models for detecting misbehavior.  For this perspective, the scope of this survey is wider. It focuses on the use of ML in the misbehavior detection systems.

\subsection{Development and Evaluation}

\begin{figure}[!ht]
		\centering
		\includegraphics[width=9cm,height=3.5cm]{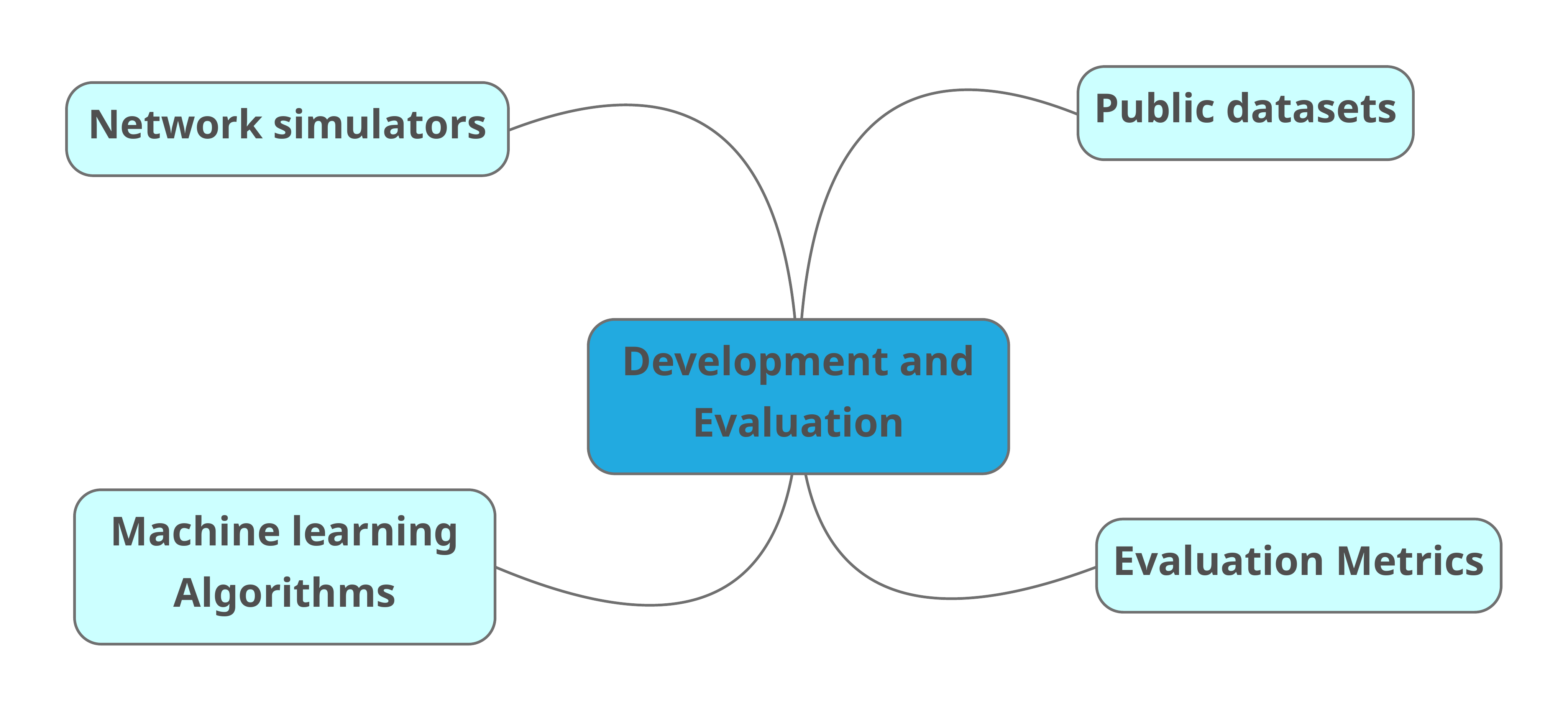}
	    \caption{Development and evaluation elements}
		\label{fig:deveval}
\end{figure}

This subsection describes different elements used to develop and evaluate ML-based MDSs. As depicted in Figure~\ref{fig:deveval}, these elements include public datasets, network simulators, and evaluation metrics. 

\subsubsection{Public datasets}

Several public security datasets were used to build ML-based MDSs. In the following, we briefly describe these datasets.

\begin{itemize}
\item \textbf{\textbf{VeReMi}}~\cite{Veremi, VeReMilink} is a simulated dataset generated using simulation tools such as OmNeT++ and Veins. Five types of position falsification were implemented. i) Constant: which consists in broadcasting fixed positions; (ii) Constant offset: which consists in broadcasting a fixed offset added to the real positions; (iii) Random: which consists of broadcasting random positions belonging to the simulated area; (iv) Random offset: which consists in broadcasting random positions that belong to a rectangle around the vehicle; and (v) Eventual stop: in which the attacker behaves normally for some time and then attacks by broadcasting a fixed position for a period. This dataset is generated for different traffic densities and different attacker radios. 
    
\item \textbf{VeReMi Extension}~\cite{kamel2020veremi, VeReMiExtendedlink} is also a simulated dataset generated using the Framework For Misbehavior Detection (F2MD)~\cite{kamel2020simulation, F2MDlink}, which is based on OmNeT++ and Veins. This dataset represents an extension of VeReMi implementing nice type of attacks: (1) Position falsification (constant, random, constant offset, and random offset); (2) Speed Malfunctions (constant, random, constant offset, and random offset); (3) Delayed Messages; (4) DoS attacks; (5) DoS Random; (6) Data Replay; (7) Disruptive; (8) Eventual Stop; (9) Traffic congestion Sybil.
    
\item \textbf{DARPA}~\cite{darpa1, darpa2}  are popular intrusion detection datasets created using an emulated network environment at the MIT Lincoln Lab. They implement attacks on authentication such as scanning attacks, User to Root (U2R), and Remote to Local (R2L) attacks. They also implement attacks on availability like Dos attacks.

\item \textbf{CAIDA DDos2007}~\cite{caida} includes approximately one hour of traffic traces from a DDoS attack (UDP flooding) attempting to block access to a server by consuming its computing and network resources. 

\item \textbf{AWID 2}~\cite{kolias2015intrusion, AWID2} implements popular attacks on 802.11 generated based on a small testbed. It includes various attacks on authentication (e.g, ARP injection), availability (e.g, RTS beacon), and confidentiality (e.g, rogue AP). 

\item \textbf{KDD CUP 99}~\cite{kdd} is a popular intrusion  detection  dataset, which includes 23  attacks on authentication and availability such as R2L, probing Attack, DoS , and U2R.

\item \textbf{NSL-KDD}~\cite{tavallaee2009detailed,NSL-KDD} came to enhance the KDD CUP 99 by removing duplication and creating more sophisticated sub-datasets. NSL-KDD includes the same attacks as the KDD CUP 99.

\item \textbf{Kyoto}~\cite{song2006description,Kyoto} is a honeypot dataset which contains real packet-based traffic converted into a new format called sessions. Each session comprises 24 features, 14 out of them are inspired from the KDD CUP 99 dataset. The remaining 10 are flow-based features such as IP addresses , ports, or duration.
    
\item \textbf{UNSW-NB15}~\cite{moustafa2015unsw, unsw} was created in a small emulated environment over 31 hours. It includes nine attacks such as backdoors, DoS, exploits, fuzzes, or worms.  

\item \textbf{CICIDS2017}~\cite{sharafaldin2018toward,CICIDS2017} was created within an emulated environment over a period of 5 days. It contains a wide range of attack types like SSH brute force, heartbleed, botnet, DoS, DDoS, web and inﬁltration attacks. 

\item  \textbf{CRAWDAD (mobiclique)} \cite{thlab-sigcomm2009-20120715} includes the traces of Bluetooth encounters, opportunistic messaging, and social profiles of 76 users of MobiClique application at SIGCOMM 2009.

\item \textbf{NGSIM trajectory datasets}~\cite{ngsim} includes longitudinal and lateral positioning information for all vehicles in certain regions. 
\end{itemize}

Table~\ref{tab:dssecurity} shows the types of security attacks provided by each of the security datasets concerning security services described in the subsection~\ref{sec:securityservices}. It is worth mentioning that any dataset includes attacks on non-repudiation. In addition, as we can see that table doesn't include CRAWDAD (Mobiclique) and NGSIM  trajectory datasets since these data are not originally included attacks but are used by authors after pre-processing like injecting noise.

\begin{table}[]
\centering
\caption{Datasets vs. Security services}
\label{tab:dssecurity}
\begin{tabular}{|l|c|c|c|c|}
\hline
Dataset          & \rotatebox{90}{~\parbox{2cm}{Integrity~}} & \rotatebox{90}{~\parbox{2cm}{Authentication~}} & \rotatebox{90}{~\parbox{2cm}{Availability~}}& \rotatebox{90}{~\parbox{2cm}{Confidentiality~}} \\ \hline
VeReMi           & X         &                &              &                 \\ \hline
VeReMi extension & X         &                & X            &                 \\ \hline
DARPA            &           & X              & X            &                 \\ \hline
CAIDA  DDos 2007 &           &                & X            &                 \\ \hline
AWID2            &           & X              & X            & X               \\ \hline
KDD CUP 99       &           & X              & X            &                 \\ \hline
NSL-KDD          &           & X              & X            &                 \\ \hline
Kyoto            &           & X              & X            &                 \\ \hline
UNSW-NB15        &           & X              & X            &                 \\ \hline
CICIDS2017       & X         & X              & X            &                 \\ \hline
\end{tabular}
\end{table}

\subsubsection{Network simulators}
Several network simulators have been used to generate customized security datasets. In the following, we describe these network simulators.

\begin{itemize}

    \item \textbf{OMNeT++ (Objective Modular Network Testbed in C++)}\footnote{https://omnetpp.org/}: is a modular, component-based C++ simulation library and framework, primarily for building network simulators. OMNeT++ itself is a simulation framework without models for network protocols like IP or HTTP. The main computer network simulation models are available in several external frameworks. 
    
     \item \textbf{SUMO (Simulation of Urban MObility)}\footnote{https://www.eclipse.org/sumo/} is a microscopic mobility simulator. It allows to build realistic traffic and mobility models of entire cities for a variety of application ares. It supports the modeling of pedestrians, bicycles, passenger cars, trucks, busses, trains and even ships.

    \item \textbf{NS2/NS3 (Network Simulator Version 2/3)}\footnote{https://www.nsnam.org/} is an open-source event-driven simulator designed specifically for research in computer communication networks.
    
    \item \textbf{Veins}\footnote{https://veins.car2x.org/} is an open source framework for running vehicular network simulations. It is based on two well-established simulators: OMNeT++ and SUMO. It extends these to offer a comprehensive suite of models for vehicular networks simulation.

    \item \textbf{CTUns-5.0} \cite{ctuns} is a high-fidelity and extensible network simulator and emulator capable of simulating various protocols used in both wired and wireless IP networks.
    .
    \item \textbf{GloMoSim (Global Mobile Information System Simulator)} \cite{zeng1998glomosim} is a network protocol simulation software that simulates wireless and wired network systems. GloMoSim supports protocols for a purely wireless networks.
    
    \item  \textbf{Mininet}\footnote{http://mininet.org/} is a network emulator which creates a network of virtual hosts, switches, controllers, and links. Mininet hosts run standard Linux network software, and its switches support OpenFlow \cite{mckeown2008openflow} for highly flexible custom routing and Software-Defined Networking.
    
\end{itemize}

\subsubsection{Machine learning algorithms}
\begin{figure*}[!ht]
		\centering
		\includegraphics[width=18cm,height=8cm]{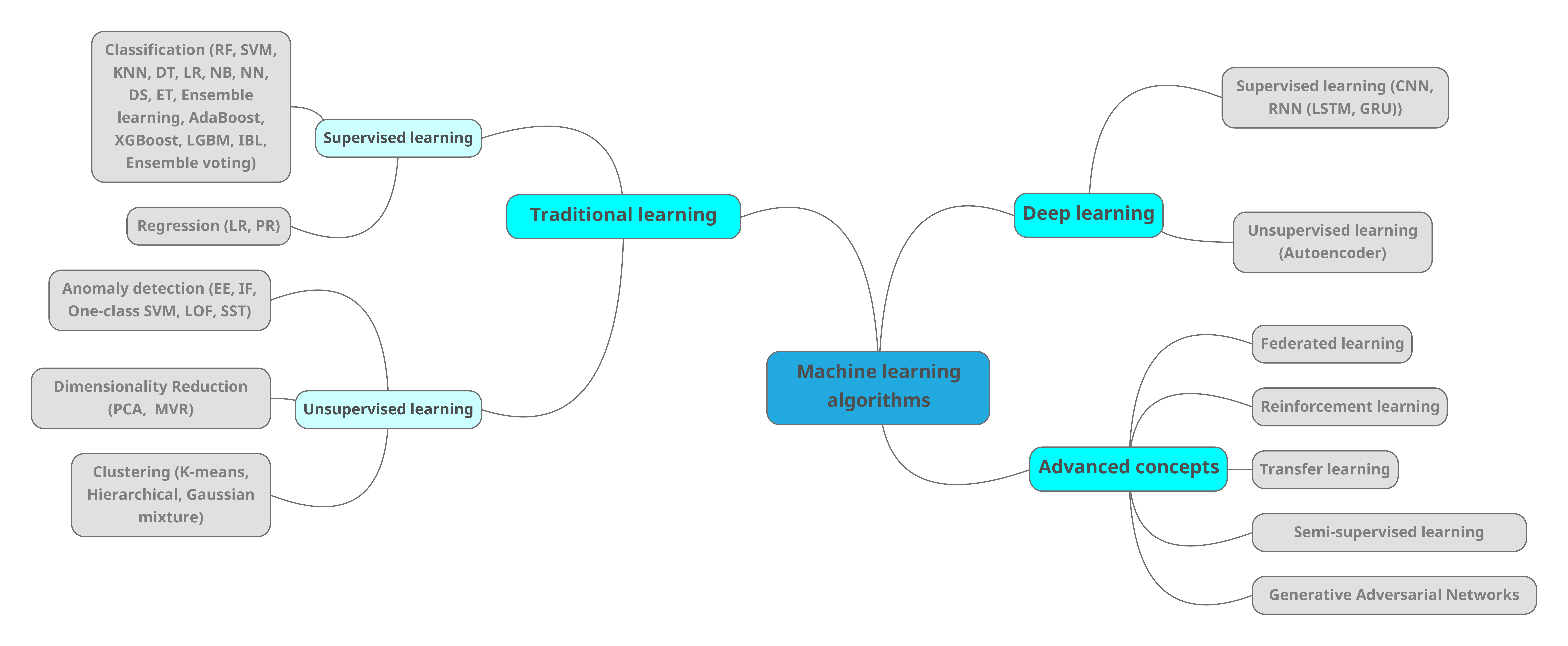}
	    \caption{Machine learning algorithms}
		\label{fig:MLalgorithms}
\end{figure*}

"Machine learning is a branch of artificial intelligence (AI) and computer science which focuses on the use of data and algorithms to imitate the way that humans learn, gradually improving its accuracy" \cite{ibmml}. This section gives a brief introduction of various machine learning techniques and concepts that is used to build ML-based MDSs. Figures~\ref{fig:MLalgorithms} summarizes these techniques and concepts. This section is divided into three subsections: traditional learning, deep learning, and advanced ML concepts.

\subsubsection*{1) Traditional learning}
Traditional learning refers to machine learning algorithms that aren't based on deep learning, as explained in the following section.

\begin{itemize}
    \item Supervised learning is an ML approach that leverages labeled datasets to train or “supervise” algorithms into classifying data or predicting outcomes. Supervised learning can be separated into two types of problems: classification and regression:
    
    \begin{itemize}
        \item  \textbf{Classification} problems use an algorithm to classify data into specific categories. For example, classifying safety messages into two groups: malicious or normal. Common classification algorithms are  Naive Bayes (NB),  Logistic Regression (LR), Support Vector Machine (SVM), K-Nearest Neighbor (KNN), Random Forest (RF), Neural Networks (NN), Extra Tree (ET), AdaBoost, Decision Stump (DS),  Ensemble learning (bagging, boosting, stacking), XGBoost, Light Gradien Boosting Machine (LGBM), Instance Based Learning (IBL), and Ensemble voting.
        \item  \textbf{Regression} problems use an algorithm to predict real or discrete input variables. For example, predicting a trust value of a V2X node or the number of attackers in the V2X network. Common regression algorithms are Linear Regression (LR) and Polynomial Regression (PR).
    \end{itemize}
    \item \textbf{Unsupervised learning} In contrast to supervised learning, unsupervised learning uses unlabeled datasets for finding patterns that help to understand data structure. Supervised learning can be classified into three types of problems: anomaly detection, clustering, and dimensionality reduction. 
    
    \begin{itemize}
        \item \textbf{Clustering} is commonly used to organize data into groups that are easier to comprehend and manage. Common clustering algorithms are k-means, hierarchical, and Gaussian mixture models.
        \item \textbf{Anomaly detection} consists in identifying unexpected items or events in the dataset without any prior knowledge. Common anomaly detection algorithms are Elliptic Envelope Algorithm, Isolation Forest Algorithm, One-class SVM Algorithm, Singular Spectrum Transformation (SST), and Local Outlier Factor (LOF) Algorithm.
        \item \textbf{Dimensionality reduction} consists in transforming data from a high-dimensional space into a low-dimensional space so while preserving some important information quantity from the original data. Common dimensionality reduction algorithms are Principal Component Analysis and Missing Value Ratio.
        
    \end{itemize}
    
\end{itemize}

\subsubsection*{2) Deep learning}
Deep learning (DL) is a subset of ML that is based on Neural Networks (NN). "Deep” refers to the number of hidden layers required to train ML models. The DL algorithm outperforms ML algorithms, especially in large data sets with a huge number of features and rows as well. DL algorithms have enabled advances in several applications such as computer vision, natural language processing, and machine translation. DL is offering efficient learning algorithms for both supervised and unsupervised tasks.

\begin{itemize}
    \item \textbf{Supervised learning:}  Convolutional Neural networks (CNN) and Recurrent Neural Networks (RNN) are the more common DL learning algorithms:
    
    \begin{itemize}
        \item \textbf{CNN:} are specialized DL algorithms designed for computer vision applications. CNN architectures take images represented as a matrix of pixels. CNN combines traditional layers in NN with more sophisticated operators such as convolution and polling operators to learn fine-tuned features from the figures.
        \item \textbf{RNN:} RNNs are DL algorithms addressing problems involving data sequences or time series such as speech recognition, natural language processing, and language translation. RNNs are inter-connecting learning nodes enabling a kind of memory that takes knowledge from previous data sequences to influence the current data sequence and the output. Several advanced RNN architectures are proposed such as Long short-term memory (LSTM) and Gated recurrent units (GRUs).
    \end{itemize}
    
\item Unsupervised learning algorithms:

\begin{itemize}
    \item \textbf{Autoencoder} is an unsupervised deep learning algorithm that uses a neural network architecture with a tiny bottleneck layer in the middle that contains the input data's encoding representation to reconstruct the input data in the output. Specifically, the autoencoder consists of (i) the encoder that compresses the data inputs to encoding presentation with a smaller size than the input and (ii) the decoder that takes the encoding representation and tries to reconstruct the input data.  In unsupervised anomaly detection, Autoencoders aim to minimize the reconstruction error as part of its training. The reconstruction loss is used to detect the anomalies. 
\end{itemize}

\end{itemize}

\subsubsection*{3) Advanced ML concepts}

\begin{itemize}

 \item \textbf{Federated learning (FL)} is a distributed ML technique enabling collaboration between multiple nodes to collaboratively build a global model without sharing their data sets. The training of the global model is performed within several rounds until the FL server achieves a satisfactory global model. In each round, the FL server sends the global model to a set of selected nodes. Each learning node uses its local labeled data set to calculate its local updates of the global model. At the end of the round, all the selected learning nodes send their local updates to the FL server. Once all the updates are received, the FL server aggregates local updates for calculating the new global model.

\item \textbf{Reinforcement learning} is a type of goal-oriented learning that trains models on how to achieve specified goals while maximizing outcomes over time. It is based on rewarding positive behaviors while penalizing those that are undesirable. Reinforcement learning agents can perceive and interpret their surroundings, as well as taking actions and learning through trial and error. 

\item \textbf{Transfer learning} is an ML technique that focuses on exploiting the knowledge gained by solving a given problem to apply it to another related problem. For example, the knowledge acquired from learning to detect DoS attacks could be used to detect DDoS attacks.

\item \textbf{Semi-supervised learning} is similar to supervised learning, but the training process combines a small amount of labeled data with a large amount of unlabeled data during training. Semi-supervised is usually used where unlabeled data is accessible, but labeled data is hard to obtain.

\item \textbf{Generative Adversarial Networks (GAN)}: GAN is a deep learning network that can create data that is similar to the input data. It consists of two networks that train together: (i) Generator: which generates data with similar characteristics as the training data, and (ii) Discriminator: which attempts to categorize observations as "real" or "created" given data comprising observations from both the training data and produced data from the generator. 
    
\end{itemize}

\subsubsection{Evaluation metrics}
Various evaluation metrics have been used to evaluate ML-based MDSs. In the following, we give the calculating formula for each metric with a short description.

\begin{itemize}

\item \textbf{Accuracy} is the ratio of the correctly detected attackers to the total of vehicles.

    \begin{equation}
    Accuracy = \frac{TP+TN}{TP+TN+FP+FN}
   \end{equation}
   \begin{itemize}
       \item \textbf{True positive (TP)} is the number of cases correctly identified the attackers. 
       \item \textbf{False positive (FP)} is the number of cases incorrectly identified the attackers.
       \item \textbf{True negative (TN)} the number of cases correctly identified identified the honest vehicles.
       \item \textbf{False negative (FN)} is the number of cases incorrectly identified the honest vehicles. 

   \end{itemize}
   
\item \textbf{Precision} calculates the ratio of correctly detected attackers to the total detected attackers. 

     \begin{equation}
     Precision = \frac{TP}{TP+FP}
     \end{equation}

\item \textbf{Recall} calculates the ratio of correctly detected attackers to the total actual attackers. 

     \begin{equation}
          Recall = \frac{TP}{TP + FN} 
     \end{equation}

\item \textbf{F1-score} can be interpreted as a weighted average of precision and recall.
    \begin{equation}
        F1-score = 2 X \frac{Precision * Recall}{Precision + Recall}
    \end{equation}
\item \textbf{True Positive Rate (TPR)}, so-called also Sensitivity, is the proportion of attackers who has a positive detection result.
    \begin{equation}
        TPR = \frac{TP}{TP+FN}
    \end{equation}
\item \textbf{True Negative Rate (TNR)} , so-called also Sensitivity, Specificity is the proportion of honest vehicles who has a negative detection result.
    \begin{equation}
        TNR = \frac{TN}{FP+TN}
    \end{equation}
\item \textbf{False Positive Rate (FPR)} is the proportion of honest vehicles who has a positive detection result.
    \begin{equation}
        FPR = \frac{FP}{FP+TN}
    \end{equation}
\item \textbf{False Negative Rate (FNR)}: is the proportion of attackers who has a negative detection result.
    \begin{equation}
        FNR = \frac{FN}{TP+FN}
    \end{equation}
    
\item \textbf{The Receiver Operator Characteristic (ROC) curve} shows the trade-off between sensitivity and specificity. Classifiers with curves that are closer to the top-left corner perform better. 

\item \textbf{The Area Under the Curve (AUC)} is used as a summary of the ROC curve. It measures the ability of a classifier to distinguish between classes. 
\end{itemize}

\section{ML-based Misbehavior Detection Systems: a taxonomy}
\label{sec:taxonomy}

In this section, we review different ML-based MDS that have been proposed in the literature. We classify the proposed ML-based MDSs into three categories: (i) ML-based MDSs for Non-IP-based (Safety) applications: that mainly detect attacks on the facilities layer; (ii) ML-based MDSs for IP-based (non-safety) applications: that mainly detect attacks on the transport and networking layers; and (iii) ML-based MDS that can be used for both: that mainly detect attacks on the physical layer.  Each category is divided into subcategories according to the attack detected by the ML-based MDS. 

\begin{figure*}[!ht]
		\centering
		\includegraphics[width=18cm,height=7cm]{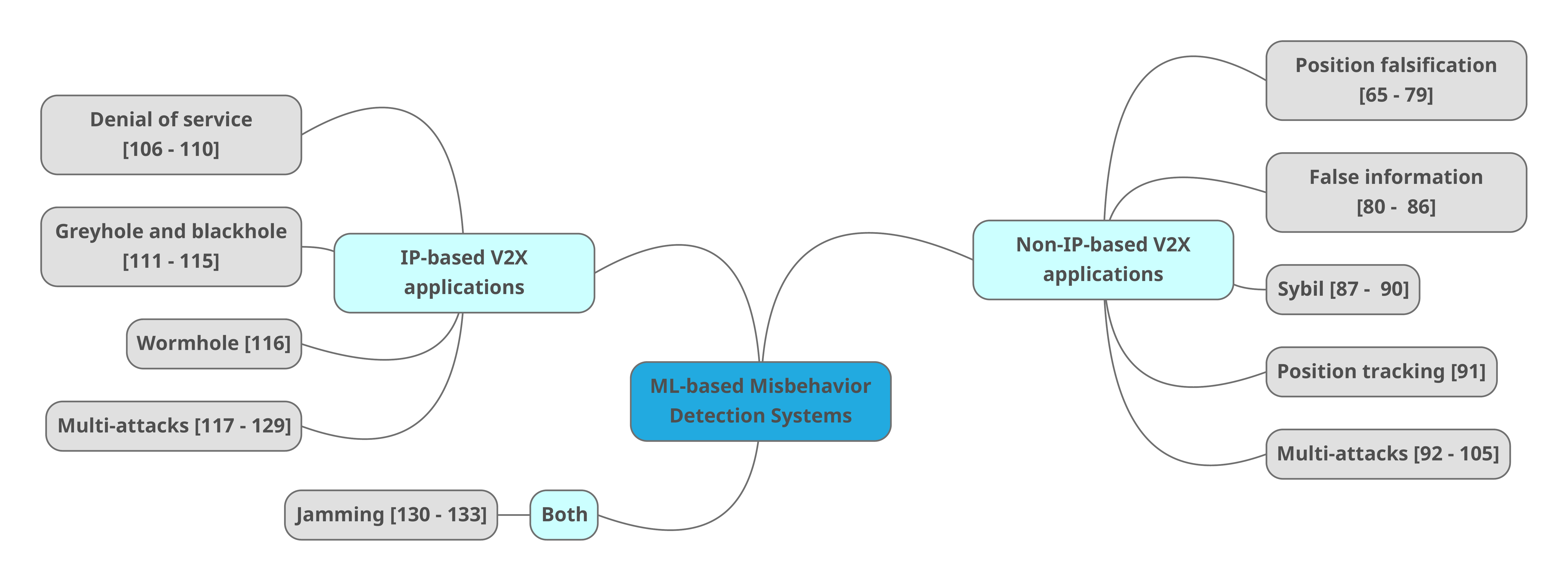}
	    \caption{Taxonomy of ML-based Misbehavior Detection Systems}
		\label{fig:gm3}
\end{figure*}

\subsection{Non-IP-based V2X applications}
This category includes ML-based MDSs detecting position falsification, false information, Sybil, position tracking, and multi-attacks. False information subcategory can comprise position falsification subcategory. However, due to the important works on position falsification, we categorized them separately. In addition, the multi-attack category includes ML-based MDSs that can detect two or more attacks.

\subsubsection{Position falsification}
 
So et al. \cite{so2018integrating} proposed an ML-based MDS that detects position falsification attacks. The proposed system built a supervised learning model based on VeReMi dataset. Six features were considered for the training: (1) Location Plausibility check; (2) Movement Plausibility check; (3) Average distance between the first received beacon and the final received beacons; (4) average velocity between the first received beacon and the final received beacon; the feature (5) is the magnitude of features 3 and 4; finally the feature (6) is the total displacement between two received messages. Two ML algorithms were evaluated SVM and KNN, and precision and recall were used for evaluation metrics. Le et al. \cite{le2019} also proposed an ML-based MDS based on supervised learning to detect position falsification attacks. The paper leverages comparing the trajectory of vehicles with trajectories of legitimated vehicles. Three features are proposed to compare the trajectories : (i) Movement Plausibility Check (MPC): which checks if their positions are not changed between two consecutive messages received from the same vehicle. (ii)  Minimum Distance to Trajectories (MDT): which measures the similarity between observed trajectories and legitimate trajectories; and (iii) Minimum Translation Distance to Trajectories (MTDT): which checks if any offset is added to received positions. Base on the proposed features, a multi-class classifier is trained on VeReMi data set to detect the five false position attacks. Two classification algorithms (SVM and KNN) were tested based on MATLAB implementation. Precision and recall were used as evaluation metrics. Singh et al. (1) \cite{singh2019machine} proposed a supervised-based  MDS to detect position falsification attacks. The training was done based on VeReMi dataset. Three combinations of features were tested (i) (position, speed) (ii) (position + $\Delta$ position (between the sender and the receiver), and (iii) (position, speed, $\Delta$ position, $\Delta$ speed). Two ML algorithms for a binary classification were tested SVM and LR. The F1-score is the metric used to evaluate the performance. Sharma et al. (1) \cite{sharma2020machine} proposed an ML-based MDS to detect position falsification attacks. The proposed system combines plausibility checks with ML models. The model was trained on VeReMi dataset based on a supervised learning approach. Four features were used for the trained position coordinates (x,y) and the speed coordinates ($v_{x}$, $v_{y}$). Six ML algorithms were compared SVM, KNN, NB, RF, Ensemble boosting, Ensemble voting.  Accuracy, precision, recall, and F1-score were used as evaluation metrics. Kosmanos et al. \cite{kosmanos2020novel} proposed a supervised approach to detect position falsification attacks. The ML model was based on a binary classification and was trained based on a dataset generated using the Veins network simulator. Four features were used for the training (i) the Signal Strength Indicator; (ii) the Signal Quantity Indicator; (iii) the Packet Delivery Ratio and (iv) PVRS, which is a position verification check based on the speed and the GPS position. Two classification algorithms were tested: KNN and RF. FPR, TPR, and ROC were used as evaluation metrics. Montenegro et al. \cite{montenegro2020detection} proposed an ML-based MDS based to detect position falsification attacks based on a supervised learning approach. Four types of position falsification attacks were implemented using Veins simulator. The trust value is the only feature is considered for the training. The trust value is calculated based on the weighted sum of the normalized speed and the normalized received power. A binary classification based on KNN was adopted for training the ML model. Four evaluation metrics were used: accuracy, recall, TPR, and the FPR. Ecran et al. (1) \cite{ercan2021new} proposed a supervised learning approach to detect position falsification attacks. The learning is based on the VeReMi dataset. Two combinations of features were considered: (i) (RSSI, position, distance between sender and receiver, $\Delta$ position of sender, the estimated angle of arrival (AoA), the estimated distance between the sender and the receiver), and (ii) (Position, the distance between sender and receiver, estimated AoA, estimated distance between sender and receiver). Two ML algorithms (binary) were tested KNN and RF. The used evaluation metrics are precision, recall, accuracy, and, F1-Score. The authors proposed an extension of this work in Ecran et al. (2) ~\cite{ercan2021misbehavior}. Unlike their previous work, multi-class classifiers were trained instead of binary classifiers. In addition, ensemble Learning, which combines KNN and RF was also considered. Hawlader et al. \cite{hawlader2021intelligent} proposed an ML-based MDS based on a supervised learning approach. The proposed models were trained on the VeReMi dataset. Twenty features were extracted based on the difference in positions sent by vehicles. Both binary and multi-class classifiers were trained. Six ML algorithms are tested: SVM, DT, RF, KNN, NB, and LR. Accuracy, precision, recall, and F1-score were used as evaluation metrics. The models were also validated using simulations. Okamura et al. \cite{okamura2021misbehavior} proposed unsupervised anomaly-based MDS that detects position falsification attacks. Four types of position falsification attacks similar to the proposed in VeReMi were implemented using the Scenargie network simulator. The position was used to detect the attack after transforming them into time series using SST. Precision, recall, and F1-score were used as evaluation metrics. The anomaly detection system was deployed on the cloud. Grover et al.(1) \cite{9395714} proposed an unsupervised anomaly detection MDS to detect position falsification attacks. The VeRemi dataset was used to train the ML models. Several ML models have been compared: GRU (1 layer), LSTM (1 layer with changing the number neurons), and stacked LSTM with changing the number of the layers from 2 to 5. The used evaluation metrics were accuracy and recall. The best results are observed using the stacked LSTM with 4 layers. The paper also proposed to deploy the ML models on the edge nodes. Sedar et al. \cite{reinforcement} proposed an ML-based MDS based on reinforcement learning to detect sudden-stop (eventual stop) attacks, which is a specific type of position falsification attack. The VeReMi extension dataset was used to train and evaluate the ML model. The proposed model was trained based on one feature either the position or the speed. The used evaluation metrics were precision, recall, and F1-score. Uprety et al. \cite{uprety2021privacy} leveraged Federated learning to propose privacy preservation collaborative ML-based MDS for position falsification attacks. This VeReMi dataset was used for the training and the evaluation. Four features were used for the training. The two first features are the difference of calculated average velocity and predicted ones in x and y directions respectively. Feature 3 is the magnitude of features 1 and 2. The last feature is the difference between calculated total displacement and the predicted total displacement. The evaluation metrics were precision and recall. Sharma et al. (2) \cite{sharma2021machine, sharma2021machine2} proposed a supervised-based MDS to detect position falsification attacks. The ML model was trained using the VeReMi dataset. The selected features for training are position and speed features from two consecutive beacons. Both binary and multi-class classifiers were trained to detect the attacks. Several ML algorithms were used to build the ML model: KNN, RF, NB, and DT. Several metrics were used for the training: Precision, Recall, and F1-score.

\subsubsection{False Information}
Ghaleb et al. \cite{ghaleb2017effective} proposed a ML-based detection system to detect data injection. The mobility traces were extracted for the NGSIM dataset and replayed in MATLAB. The data injection attack was implemented by injecting dynamic noise where 20\% of vehicles are considered malicious. The ML model was trained using seven features: (1) Overlaying check; (2) consistency of reported uncertainties; (3) mobility message prediction error; (4) communication-based feature; (5) Appearance position-based features; (6) average Mobility messages prediction error, (7) the time to last received mobility message. The paper adopts a supervised approach based on NN. Accuracy, F1-score, recall, and precision were used as evaluation metrics. Monteuuis et al. \cite{monteuuis2018my} proposed supervised learning-based MDS to detect the absence of correlation between the type V2X entity and the dimension of the vehicles. For training and evaluation, open datasets about cars, motorcycles, and pedestrians were collected from the internet and processed (cleaned and the number of features was reduced).  Three features were selected: the width, length, and type of V2X entity. Misbehaviors were generated by injecting noise into the dataset. Three ML algorithms were tested: NN, AdaBoost and RF. Several metrics were used for the evaluation:  TPR, TNR, FPR, FNR, accuracy, and F1-score. Singh et al. (2) \cite{singh2018misbehavior} proposed an ML-based MDS for detecting malicious infrastructure nodes reporting false information to the traffic management center. The evaluation dataset was generated using the SUMO mobility generator. The proposed system was based on predictive models built using a neural network and LSTM algorithm to predict the traffic congestion (the halt time in traffic segment) based on data provided by loop detectors. To detect the attack, the information reported by the infrastructure nodes was compared with the output of the proposed model. The system is installed in the traffic management center.  Gyawali et al. \cite{gyawali2019, gyawali2020} proposed an ML-based MDS for detecting false alert and position falsification attacks. For detecting a false alert attack, a binary classifier was trained based on the difference between the average flow value and the received flow value from the vehicles. The flow value is calculated based on the density of vehicles and the average speed of vehicles. The dataset for this attack was generated using the Veins simulation platform. On the other hand, for detecting a false position attack, a multi-class classifier was trained based on the VeReMi dataset. The considered features were the change in speed and position between two consecutive beacons, the receiving distance, the RSSI, the change in its speed and position. Several classification algorithms were used in this paper: LR, KNN, DT, Bagging, RF. Precision, recall, F1-score were used as evaluation metrics. Negi et al. \cite{negi2020} proposed an anomaly detection-based MDS. The proposed system leverages an unsupervised learning approach based on LSTM. The datasets were generated from experiments performed on a treadmill-based autonomous car simulator at the University of Waterloo. This system focus on detecting anomalies in big data generated by connected vehicles rather than focusing on V2X attacks. To speed up the training process, a cluster of servers instead of one server. Once the anomaly detection model was trained, the parameters of this model were distributed to vehicles for the real-time detection of anomalies. To keep the model updated, the model was retrained over time based on data newly collected, and the parameters of the new model are distributed to vehicles. AUC was used as an evaluation metric. Almalki et al. \cite{almalki2021deep} proposed a supervised-based MDS detect false data inject attacks. The authors proposed to take several contextual data in addition to data collected in real-time for attack detection. The authors used the NGSIM dataset, which contains data acquired from the environment using a set of sensors. In this MDS, data undergo several pre-processing steps including missing values imputation base don the local and global fuzzy-clustering correlation approach. The ML models were trained using LR, SVM, and CNN. Accuracy, F1-score, Detection Rate (DR), and FPR were used as evaluation metrics.

Table~\ref{tab:MDSfalseinfo} summarizes the type of false information detected by previously described ML-based MDSs. There are four types of false information: dimension and type of vehicles, position, alert, and road traffic. We can see that three of the described ML-based MDSs don't explicitly specify the type of detected false information.

\begin{table}[]
\caption{ML-based MDS vs. false information}
\label{tab:MDSfalseinfo}
\resizebox{9cm}{!}{
\begin{tabular}{|c|ccccc|}
\hline
                                                                & \multicolumn{5}{c|}{False information}                                                                                                                  \\ \hline
                                                                & \multicolumn{1}{c|}{\rotatebox{90}{~\parbox{1.7cm}{Not specified~}}} & \multicolumn{1}{c|}{\rotatebox{90}{~\parbox{1.7cm}{Dimension and type~}}} & \multicolumn{1}{c|}{\rotatebox{90}{~\parbox{1.7cm}{Position~}}} & \multicolumn{1}{c|}{\rotatebox{90}{~\parbox{1.7cm}{Alert~}}} & \rotatebox{90}{~\parbox{1.7cm}{Road traffic~}} \\ \hline
Ghaleb et al. \cite{ghaleb2017effective}       & \multicolumn{1}{c|}{X}             & \multicolumn{1}{l|}{}                  & \multicolumn{1}{l|}{}         & \multicolumn{1}{l|}{}      &              \\ \hline
Monteuuis et al. \cite{monteuuis2018my}        & \multicolumn{1}{l|}{}              & \multicolumn{1}{c|}{X}                 & \multicolumn{1}{l|}{}         & \multicolumn{1}{l|}{}      &              \\ \hline
Singh et al. (2) \cite{singh2018misbehavior}       & \multicolumn{1}{l|}{}              & \multicolumn{1}{l|}{}                  & \multicolumn{1}{l|}{}         & \multicolumn{1}{l|}{}      & X            \\ \hline
Gyawali et al. \cite{gyawali2019, gyawali2020} & \multicolumn{1}{l|}{}              & \multicolumn{1}{l|}{}                  & \multicolumn{1}{c|}{X}        & \multicolumn{1}{c|}{X}     &              \\ \hline
Negi et al. \cite{negi2020}                    & \multicolumn{1}{c|}{X}             & \multicolumn{1}{l|}{}                  & \multicolumn{1}{l|}{}         & \multicolumn{1}{l|}{}      &              \\ \hline
Almalki et al. \cite{almalki2021deep}          & \multicolumn{1}{c|}{X}             & \multicolumn{1}{l|}{}                  & \multicolumn{1}{l|}{}         & \multicolumn{1}{l|}{}      &              \\ \hline
\end{tabular}
}
\end{table}

\subsubsection{Sybil}

Gu et al. \cite{gu2017support} proposed a supervised learning-based MDS to detect Sybil attacks. The dataset was generated using the SUMO mobility simulator. The driving patterns of vehicles were represented as matrices. Each line of a matrix contains five fields: time, location, velocity, acceleration, acceleration variation at time t. The number of lines is the period for which this matrix is constructed. The two max eigenvalues were used as training features. The SVM and the neural network's algorithms were used for binary classification. The used evaluation metrics were TPR, FPR, and FNR. The same authors in \cite{gu2017k} also proposed a similar approach but used KNN for binary classification and the accuracy was used as an evaluation metric. Kamel et al. (1) \cite{kamel2019misbehavior} proposed an ML-based MDS for detecting Sybil attacks. The paper defines four types of Sybil attacks: (i) Traffic Congestion Sybil, (ii) Data reply Sybil; (iii) Dos Random Sybil, (iv) Dos Disruptive Sybil. The detection system was divided into two systems: Local and global. The location system was deployed at the vehicle level where a set of plausibility and consistency checks to detect the misbehavior and report it to the global system. The global system was equipped with ML-based MDS and is located in the cloud. The system was trained on the VeReMi extension dataset. Thirty features were used for a muti-class classification of attack types using LSTM. Accuracy, F1-score, recall, precision were used as evaluation metrics. Quevedo et al. \cite{quevedo2020} proposed an ML-based MDS for detecting Sybil attacks. A supervised learning approach was also adopted for training ML models. The training and detection of the attacks were performed on edge nodes. The collected data consists of a set of matrices describing the driving patterns of vehicles. The columns of matrices are the features considered for learning. Each row of a matrix contains driving information of a vehicle at time \textit{t}. An unsupervised learning data dimensionality reduction technique was used to reduce the dimensionality of matrices. The Extreme Machine Learning (EML) was used for classification based on the data set generated using SUMO mobility generator contains between 1\% and 20\% Sybil attackers. The used metric was accuracy.

\subsubsection{Position tracking}

Boualouache et al. \cite{boualouache2021federated} proposed an ML-based MDS for detecting position tracking attacks. The authors first identified a set of strategies that can be used by attackers to efficiently track vehicles without being visually detected. Based on these strategies, a syntactic dataset was generated to train the ML models. The MDS enables federated learning (FL) at the edge to ensure collaborative learning while preserving the privacy of vehicles. FL clients use a semi-supervised learning approach for self-labeling. Six traditional ML algorithms were used to train binary and multi-class classifiers: LR, KNN, SVM, NB, DT, and RF. In addition, a deep learning model was used for binary classification based on an FL architecture. Four evaluation metrics were used: precision, recall, F1-score, and accuracy.

\subsubsection{Multi-attacks}

Grover et al. (2) \cite{grover2011machine} proposed a supervised ML-based MDS to detect six attacks: (i) Impersonation, (ii) False position, (iii) Combination of impersonation and false position, (iv) Grey hole, (v) reply, (vi) timing. The dataset was generated using the NCTUns-5.0 network simulator. Several features considered for training: (i) Position, (ii) acceptance range, (iii) speed deviation, (iv) RSS, (v) packet transmitted, (6) packet delivery ratio, (7) packet drop ratio, (8) packet capture ratio, (9) packet capture ratio, (10) packet collision ratio, and (11) packet retransmission error ratio. Both binary classifiers and multi-classifiers were built. Five algorithms were compared NB, IBK, RF, DT, Adaboost. The metrics are TPR, FPR, TNR, and FNR. Grover et al. (3)~\cite{grover2011misbehavior} also proposed a similar ML-based MDS in~\cite{grover2011machine}. Their results demonstrated that the Ensemble-based learning gives better results than in~\cite{grover2011machine} in the case of binary classification. Li et al.~\cite{li2015svm} (1) proposed a supervised learning-based MDS detect packet dropping, packet modification, RTS (Request to Send) flooding attacks. The dataset was generated using the GloMoSim network simulator. Three features were used to train the ML model: packet drop rate (PDR), packet modification rate (PMR), and RTS flooding rate (RTS). The authors also considered using other contextual information such as velocity, channel status, temperature and wind speed, and GPS coordinate and altitude. But this information is not used in the evaluation. The SVM algorithm was used for binary classification. The used evaluation metrics were precision and recall. Zhang et al. (1) \cite{zhang2018misbehavior} proposed an ML-based MDS for detecting false messages and message suppression. The data set was generated from simulations using the Veins simulation framework. To detect false message attacks, five features were for training and evaluation were considered: (1) VehicleType, (2) MessageType, (3) reputation, distance to vehicle,  message forwarding status, type of forwarding vehicle, the reputation of and (4) the forwarding vehicle. To detect suppression message attacks, four features were considered Packet Drop Rate (PDR), Packet Delay Forward Rate (PDFR), Packet Modify Rate (PMOR), and Packet Misroute rate (PMIR). Two binary classifiers were trained and deployed on connected vehicles; one for each attack. SVM was used for training and three evaluation metrics were considered: TPR, FPR, and accuracy. Eziama et al. \cite{eziama2018malicious} proposed an unsupervised-based MDS based on Bayesian Neural Network that combines deep learning with probabilistic modeling. The paper also described three attacks: timing attack, Sybil attack, and False Position attack. However, the proposed model was not evaluated. Mahmoudi et al. \cite{mahmoudi2019towards} proposed a supervised-based MDS approach to detect attacks multiple attacks defined in the VeRiMi extension dataset. Several features were used for the training including local detection, Kinematic data, and generic features. The multi-class classifier was developed using five ML algorithms: RF, XGboost, LightGBM, NN, and LSTM. Precision, recall, and F1-score were used for the evaluation. Kamel et al. (2) \cite{kamel2019misbehavior2} also proposed a supervised-based MDS to detect multiple attacks considered in VeReMi extended dataset. The features considered for learning are almost similar to the previous work. Six models from three ML algorithms were tested SVM, NN, and LSTM. The evaluation metrics are recall, precision, F1-score, and accuracy.  Alladi et al. (1) \cite{alladi2021securing} proposed a supervised-based to detect multiple attacks considered in VeReMi extension dataset. The messages of each vehicle were used to generate sequences of 20 messages with 7 data fields: position(X,Y), velocity ($V_{X}$,$V_{Y}$), timestamp, pseudo-id, and label. Two multi-class classifiers were proposed in this paper. The first one considers two classes: normal and attacks. The second one considered position falsification attacks as faults; thereby three classes: normal, faults, and attacks. Two deep learning architectures were used: stacked LSTM and CNN-LSTM. Different models based on these two architectures were selected. The evaluation metrics were the accuracy, precision, recall, and F1-score. This work was extended in Alladi et al. (2) \cite{alladi2021artificial}. The authors proposed a similar supervised-based MDS that can be deployed as detection engines in the MEC. Two detection methods were considered: (1) Sequence Classification and (2) Sequence-image classification. For sequence classification, two models are used (i) four stacked layers LSTM and (ii) CNN-LSTM. For sequence-image classification two models also CNN and MLP. Alladi et al. (3) \cite{alladi2021deep} proposed an ML-based MDS to detect the same attacks considered in their previous works but based on an unsupervised learning approach. The ML models were trained based on normal data using the VeReMi extension data set. The considered deep learning architectures are similar to auto-encoders that take sequences of 20 messages as input, encode\&decode them to reconstruct the normal traces. The anomaly detection threshold was adjusted according to the accuracy to reconstruct the normal trace. Two different models were considered. Model1 (CNN-LSTM) and Model 2 (stacked 4-layer LSTM). The used evaluation metrics were precision, recall, F1-score, and accuracy. This work was extended in Alladi et al (4) \cite{alladi2021deepadv} with the consideration of more models for enhancing the detection performance. Kushardianto et al. ~\cite{kushardianto20212} proposed a supervised-based MDS to detect multiple attacks. Position and velocity features were used to train two ML models based on the VeReMi extension dataset. The first model was a binary classifier used to detect the attacks without specifying the type. The second model is a multi-class classifier used to determine the type of attack. Four ML algorithms were used to train these models: RF, LSM, GRU, and Deep Belief Network. The accuracy was used as an evaluation metric. Gonçalves et al. \cite{gonccalves2021intelligent} proposed an ML-based MDS to detect DoS and false information (speed, acceleration, and heading) attacks. Several multi-class classifiers were trained based on a dataset generated in their previous work~\cite{gonccalves2020synthesizing}. Several features were used for the training including position, speed, and heading. The proposed system has a hierarchical architecture with different four levels. The first level (vehicle) deploys a Decision Stump classifier. The second level is to forward messages from vehicles to RSUs. The third level (RSU) deploys an RF classifier. Finally, the fourth level (cloud) deployed an ensemble classifier that combines NN, DT, and RF. The accuracy, TPR, and TPR were used as evaluation metrics.

Table~\ref{tab:MDSvsmulatknoIP} summarizes the detected attacks of each of the previously described ML-based MDSs. These attacks include position falsification, false information, DoS/DDoS, Sybil, Replay, Timing, Greyhole/blackhole, and impersonation.

\begin{table}[]
\caption{ML-based MDSs vs. multi-attacks (non-IP-based applications)}
\label{tab:MDSvsmulatknoIP}

\resizebox{9cm}{!}{
\begin{tabular}{|c|c|c|c|c|c|c|c|c|}
\hline
                                                                     & \rotatebox{90}{~\parbox{2cm}{Position falsification~}} & \rotatebox{90}{~\parbox{2cm}{False information~}} & \rotatebox{90}{~\parbox{2cm}{DoS/DDoS~}} & \rotatebox{90}{~\parbox{2cm}{Sybil~}} & \rotatebox{90}{~\parbox{2cm}{Replay~}} & \rotatebox{90}{~\parbox{2cm}{Timing~}} & \rotatebox{90}{~\parbox{2.6cm}{Greyhole/blackhole~}} & \rotatebox{90}{~\parbox{2cm}{Impersonation~}} \\ \hline
Grover et al. (2) \cite{grover2011machine}          & X                      &                   &          &       & X      & X      & X                  & X             \\ \hline
Grover et al. (3)$\sim$\cite{grover2011misbehavior} & X                      &                   &          &       & X      & X      & X                  & X             \\ \hline
Li et al. (1) \cite{li2015svm}                      &                        & X                 & X        &       &        & X      & X                  &               \\ \hline
Zhang et al. (1) \cite{zhang2018misbehavior}            & X                      &                   &          &       &        &        & X                  &               \\ \hline
Eziama et al. \cite{eziama2018malicious}            & X                      &                   &          & X     &        & X      &                    &               \\ \hline
Mahmoudi et al. \cite{mahmoudi2019towards}          & X                      & X                 & X        & X     & X      & X      &                    &               \\ \hline
Kamel et al. (2) \cite{kamel2019misbehavior2}       & X                      & X                 & X        & X     & X      & X      &                    &               \\ \hline
Alladi et al. (1) \cite{alladi2021securing}         & X                      & X                 & X        & X     & X      & X      &                    &               \\ \hline
Alladi et al. (2) \cite{alladi2021deep}             & X                      & X                 & X        & X     & X      & X      &                    &               \\ \hline
Alladi et al. (3) \cite{alladi2021artificial}       & X                      & X                 & X        & X     & X      & X      &                    &               \\ \hline
Alladi et al (4) \cite{alladi2021deepadv}           & X                      & X                 & X        & X     & X      & X      &                    &               \\ \hline
Kushardianto et al. \cite{kushardianto20212}        & X                      & X                 & X        & X     & X      & X      &                    &               \\ \hline
Gonçalves et al. \cite{gonccalves2021intelligent}   &                        & X                 & X        &       &        &        &                    &               \\ \hline
\end{tabular}
}
\end{table}

\subsection{IP-based V2X applications}

This category includes ML-based MDSs detecting denial of Service, greyhole/blackhole, Sybil, Wormhole, and multi-attacks. The multi-attack category includes ML-based MDSs that can detect two or more attacks.

\subsubsection{Denial of Service}

Tan et al. \cite{tan2018} proposed an unsupervised learning-based MDS for detecting DoS attacks. In this paper, RSUs collect the traffic flow of vehicles, which is defined as a sequence of packets from the source to the destination. Each flow contains $\sigma$ packets with the corresponding time series. The Agglomerate Hierarchical clustering was applied for creating clusters of similar traffic flows. In each step of the clustering algorithm, the dynamic time wrapping distance \cite{dtw} is used to calculate the distance between the time series of different traffic follow. The model was built based on a dataset generated using Python. DR was used as an evaluation metric.  Singh et al. (3) \cite{singh2018} proposed a supervised learning-based MDS to detect DDoS attacks in software-defined V2X. A binary classifier was trained based on a dataset generated using the Mininet simulation tool. Several ML algorithms are tested: LR, DT, NB, SVM, KNN, NN, and Gradient boosting. Four evaluation metrics were used: TP, TN, FP, and FN. Yu et al. \cite{yu2018} also proposed an ML-based MDS for detecting DDoS attacks in software-defined vehicular networks. The paper proposed the integration of OpenFlow \cite{mckeown2008openflow} with vehicular networks with the focus on DDos. A set of features are extracted from an open flow table to use for training a set of binary-classifiers based on TCP, UDP, and ICMP network protocols. The SVM classification algorithm was used for training based on well-known datasets including DARPA, CAIDA, and DDos2007. DR was used as an evaluation metric. Sharshembiev et all~\cite{sharshembiev2021protocol} proposed unsupervised learning-based MDS to detect DoS attacks. The dataset was generated using the Veins simulation platform and the entropy-based anomaly detection technique was used to detect attacks based on the generated network flows. Precision, recall, and F1-score were used as evaluation metrics.

\subsubsection{Greyhole and blackhole}

Gruebler et al. \cite{gruebler2015intrusion} proposed a supervised learning-based MDS to detect blackhole attacks. The dataset was generated based on NS2 and SUMO where 15 features are selected for learning such as payload size, type, IP Source and Destination, and the sequence number. A binary-class classifier was trained based NN algorithm. TP, TN, FP, and FN were used as evaluation metrics. Alheeti et al. (1) \cite{alheeti2015}  proposed a supervised learning-based MDS to detect Grey hole attacks. The datasets were generated by simulating Grey hole with an adapted version of AODV protocol on the ns2 simulator and SUMO. Fifteen features were selected from basic and AODV traces to train two binary classifiers. The feature fuzzification was also performed before the training. Two classification algorithms were used for training: SVM and NN. Accuracy, TP, TN, FN, and FP were used as evaluation metrics. Zeng et al. \cite{zeng2018senior2local} proposed a multi-level ML-based MDS for detecting greyhole/black attacks. The proposed system was based on two binary classification models. The first binary classifier was an NN deployed on the RSU while the second is an SVM deployed on Clusterhead. The dataset was generated using GlobMoSim. The feature extraction was based on the work of~\cite{gruebler2015intrusion}. Accuracy and DR were used as evaluation metrics. Siddiqui et al. \cite{siddiqui2019machine} proposed a hybrid ML-based MDS that combines unsupervised and supervised learning for detecting grey hole attacks. The authors used CRAWDAD (mobiclique) dataset to extract three features: similarity, familiarity, and Packet PDR. Once the dataset was prepossessed, the unsupervised technique was applied to label that data. After that, a binary classifier was trained using two classification algorithms KNN and SVM. Accuracy was used as an evaluation metric. Acharya and Oluoch \cite{acharya2021dual} proposed a supervised learning-based MDS to detect blockhole attacks. The data was generated using based on a modified version of AODV implemented on the NS3 simulator. Seven features were considered source IP address, source, and destination port, timeFirstRxpaket, timeLastRxPacket, lost packets, and throughput. A binary classifier was trained using five ML algorithms: NB, LR, KNN, SVM, and gradient boosting. The used metrics for evaluation were the recall, precision, F1-score, accuracy, FPR and FNR, and ROC\_AUC score. 

\subsubsection{Wormhole}

Singh et al. (4) \cite{singh2019machine2} proposed a supervised learning-based to detect Wormhole attacks. The dataset was generated using the NS3 simulator. Several features were considered: source and destination IP Address, transmitted and received Bytes, dropped Bytes, FirstRxBytesTime, FirstTxBytesTime, and throughput. Two ML algorithms were used for training a binary classifier: KNN and SVM. TP, TN, FP, and FN were used as evaluation metrics.

\subsubsection{Multi-Attacks}

Alheet et al. (2) \cite{alheeti2016hybrid,ali2018intelligent}  proposed a supervised learning-based to detect network-level attacks. The training was done using the Kyoto dataset, which includes different types of attacks: networks, SQL, TCP, Malware, shellcodes, and exploit codes. The authors also proposed a technique to reduce the number of features. A multi-class classifier was trained to detect three classes: normal, known, and unknown attacks using the NN algorithm. TP, TN, FN, and FP. Kim et al. \cite{kim2017} proposed an ML-based MDS for software-defined vehicular networks where vehicles analyze the incoming traffic and forward some selected data flows to the SDN controller. Based on these data flows, the SDN controller trains a multi-class classifier based on KDD dataset using the SVM classification algorithm. Six features were considered for the training PDR, PMR (Packet modified ratio), RTS flooding rate, the channel status, the packet interval, the average packet interval in the flow, and the packet size. The parameters of the trained model are forwarded to vehicles to be used in the detection of misbehaving vehicles. Accuracy, precision, and recall were used as evaluation metrics. 
Zhang et al. (2) \cite{zhang2018distributed} proposed an ML-based MDS based on a distributed ML approach. The authors assumed that each vehicle has its own labeled data. The learned model was collaboratively built between vehicles without exchanging data sets between them. Instead of sharing the data sets, the vehicles share only updates of the loss functions. To prevent privacy leakage, the authors proposed a dual variable perturbation to provide dynamic differential privacy. A binary classifier was trained using LR based NSL-KDD dataset. The output of the loss function was used as an evaluation metric. Ghaleb et al. \cite{a2020misbehavior} proposed a collaborative ML-based MDS for multi IP-based attacks. The proposed system consists of four phases. In the first phase (Individual IDS Construction), each vehicle builds its local model based on its collected data. In the second phase (Neighboring Classifiers and Metadata Exchanging) vehicles share models according to the requests received from the neighbors. In the third phase  (Neighboring Misbehavior Evaluation) vehicles evaluate the received models to detect malicious models (nodes). In the last phase (Collaborative IDS Construction), the collaborative model is constructed based on the valid model checked in the third phase. To build the model the authors used the NSL-KDD dataset. The binary classification was used to build the model. Three classification algorithms were used RF, XGBoost, and SVM. The used classification metrics were: accuracy, precision, recall, F1 score, FPR, and FNR. Ashraf et al. \cite{ashraf2020novel} proposed unsupervised learning-based MDS for detecting multiple network attacks. The model was trained based on the UNSW-NB15 database, which includes exploits attacks, generic attacks, DoS attacks, Fuzzer attacks, and Recon attacks. A statistical method was used to extract the features. The proposed MDS was based on LSTM autoencoder architecture. Several metrics were used for the evaluation including precision, recall, accuracy, F1-score, and TPR. Shu et al. \cite{shu2020collaborative} proposed a collaborative ML-based MDS based on Software Defined Networking (SDN). The proposed MDS utilized deep learning with generative adversarial networks to enable multiple distributed SDN controllers to jointly train the ML model for the entire network. A binary classifier was trained based on the KDD99 dataset, where several evaluation metrics were used for the evaluation: accuracy, precision, recall, and  F1-score. Li et al. (2) \cite{li2020transfer}  proposed an ML-based MDS to detect multiple attacks including false information, DoS, and impersonation. The proposed MDS was based on a transfer learning approach to transfer the knowledge acquired by building ML models using a large number of labeled data of well-known attacks to detect new attacks with a small amount of labeled data. Two approaches were used to update the ML model based on the transfer learning approach: Cloud assisted approach where the unlabeled data was sent to the cloud. The task of the cloud is thus to label the data, update the model and send it back to vehicles. In the second approach, the update of the model is locally done on vehicles where the pre-training model was used to assign pseudo-labels to data. The model is then locally updated based on the pseudo-label data using the transfer learning approach. A multi-class classifier was trained on the AWID public data. The experiments show how to exploit knowledge built from detecting injection and impersonation attackers for detecting flooding attackers. Several evaluation metrics are SVM, RF, Accuracy, and FN. Bangui et al. \cite{bangui2021hybrid} proposed a hybrid approach to detect network-level. The proposed MDS combines a binary multi-classifier model to detect know attacks and unsupervised learning to detect unknown attacks. The model was trained based on the CICIDS2017 dataset using RF and a variation of Kmeans. Two evaluation metrics were used F1-score and accuracy. Yang et al. \cite{yang2021mth} proposed a multi-tiered hybrid intrusion detection system (MTH-IDS). The proposed ML-based MDS uses multi-class classification models to detect known attacks based on the CICIDS2017 dataset and the unsupervised anomaly detection models to detect unknown attacks. Several classification algorithms were used DT, RF, ET, XGBoost, a stacking ensemble model, and a Bayesian optimization with tree Parzen estimator (BO-TPE) to optimize the classification. The used anomaly-detection systems are cluster labeling and two biased classifiers, and a Bayesian optimization with Gaussian process (BO-GP) method for unsupervised learner optimization. The used metrics include accuracy, DR, FPR, and F1-score.  Khan et al. \cite{khan2021enhanced} proposed an unsupervised anomaly detection system to detect IP-based attacks such as DoS, reconnaissance, exploits, fuzzes, and generic attacks. The system was based on two stages. In the first stage, two models were proposed based on the standard state-based method. The second stage was based on the Bidirectional LSTM-Based. The models were deployed on the gateway to the connected vehicle. UNSWNB15 data set were used in the evaluation. Accuracy, recall,  precision,  and F1-score. Liu et al. \cite{liu2021blockchain} combined blockchain and FL for collaborative FL-based MDS. RSUs select FL workers from vehicles under their coverage and train global models. In addition, the blockchain system, which consists of RSUs, stores global models obtained after running consensus processes that combine the Proof-of-Work (PoW) and the Proof-of-Accuracy (PoA) algorithms to select the miner of the block(s). The binary classifier is built on the KDDCup99 dataset using Deep learning. Accuracy, precision, and recall were used as evaluation metrics. Rahal et al.~\cite{rahal2022antibotv} proposed supervised learning-based MDS to detect DoS and eavesdropping attacks.  A multi-class classifier was trained based on a data set generated using NS3. Several ML algorithms were used to build this classifier: KNN, NN, SVM, RF, DT, and the NB. Precision, recall, F1-score, accuracy, FPR, and FNR were used as evaluation metrics.

Table~\ref{tab:MDSvsmulatkIP} summarizes the detected attacks of each of the previously described ML-based MDSs. These attacks include impersonation, DoS/DDoS, false information, and eavesdropping. Almost all works can detect impersonation and DoS/DDoS attacks. 

\begin{table}[]
\label{}
\caption{ML-based MDSs vs. multi-attacks (IP-based applications)}
\label{tab:MDSvsmulatkIP}
\resizebox{9cm}{!}{
\begin{tabular}{|l|c|c|c|c|}
\hline
                                                                             & \rotatebox{90}{~\parbox{2cm}{Impersonation~}} & \rotatebox{90}{~\parbox{2cm}{DoS /DDoS~}} & \rotatebox{90}{~\parbox{2cm}{False information~}} & \rotatebox{90}{~\parbox{2cm}{Eavesdropping~}} \\ \hline
Alheet et al. (2)  \cite{alheeti2016hybrid,ali2018intelligent}  & X             & X         &                   &                 \\ \hline
Kim et al.   \cite{kim2017}                                 & X             & X         &                   &                 \\ \hline
Zhang et al. (2)   \cite{zhang2018distributed}                         & X             & X         &                   &                 \\ \hline
Ghaleb et al. \cite{a2020misbehavior}                       & X             & X         &                   &                 \\ \hline
Ashraf et al.   \cite{ashraf2020novel}                      & X             & X         &                   &                 \\ \hline
Shu et al.   \cite{shu2020collaborative}                    & X             & X         &                   &                 \\ \hline
Li et al. (2)  \cite{li2020transfer}                            & X             & X         & X                 &                 \\ \hline
Bangui et al.   \cite{bangui2021hybrid}                     & X             & X         &                   &                 \\ \hline
Yang et al.   \cite{yang2021mth}                            & X             & X         &                   &                 \\ \hline
Khan et al.   \cite{khan2021enhanced}                       & X             & X         &                   &                 \\ \hline
Liu et al.   \cite{liu2021blockchain}                       & X             & X         &                   &                 \\ \hline
Rahal et al. \cite{rahal2022antibotv}                  &               & X         &                   & X               \\ \hline

\end{tabular}
}
\end{table}

\subsection{Both}

This category only includes jamming attacks. 

\subsection{Jamming}

Karagiannis et al. \cite{karagiannis2018} proposed unsupervised learning-based MDS distinguish between intentional interference (jamming) and unintentional interference. Several features were considered for learning including Received Signal Strength and interference (RSSI), Packet Delivery Ratio (PDR),  Signal to Noise and Interference Ratio (SNIR), and relative Speed Variation (RSV). The k-means clustering algorithm is used in the unsupervised learning process. The dataset is generated using R programming language considering a scenario that considers interference and different types of radio jammers (constant and smart). The purpose of the evaluation is to evaluate the significance of the proposed metric to differentiate between interference and jamming and to identify certain characteristics of the jamming case. Lyamin et al. \cite{lyamin2018ai} proposed unsupervised learning-based MDS for jamming attacks. Two jamming attacks were considered: (i) Random jamming: each transmitted CAM is jammed independently with a probability $p$ and (2) ON-OFF jamming: in the OFF state no packets are jammed, while in the ON state K subsequent CAMs are destroyed with probability 1. The dataset was generated by simulations in MATLAB. The proposed approach is combined with a previous method from the same authors \cite{lyamin2013real} to propose a hybrid method to enhance the results. The used evaluation metrics are the F1-score, true positive rate (TPR), and true negative rate (TNR), while the inference is done in the connected vehicles. Abhishek and Gurusamy~\cite{abhishek2021jade} proposed an unsupervised learning-based MDS to jamming attacks. The dataset was generated using the NS3 simulator. Two features were used for the training: Packet drop ratio and Inverse packet delivery ratio. The one-class SVM algorithm was used for the training. The detection probability was used as a metric.  

\section{Summary \& Discussion}
\label{sec:summary}
In this section, we provide summaries and discussions to get an overview of the presented ML-based MDSs. This section is divided into two parts. security and privacy-oriented summary and ML-oriented summary. In the security and privacy-oriented summary, we mainly focus on detected attacks, the general design, and other different security-related aspects.  In the ML-oriented summary, we mainly focus on the ML model used to detect the attack. We analyze ML model characteristics such as the data sets, used ML algorithms, and metrics. 

\subsection{Security and Privacy oriented summary}
\begin{figure*}[!ht]
		\centering
		\includegraphics[width=18cm,height=7cm]{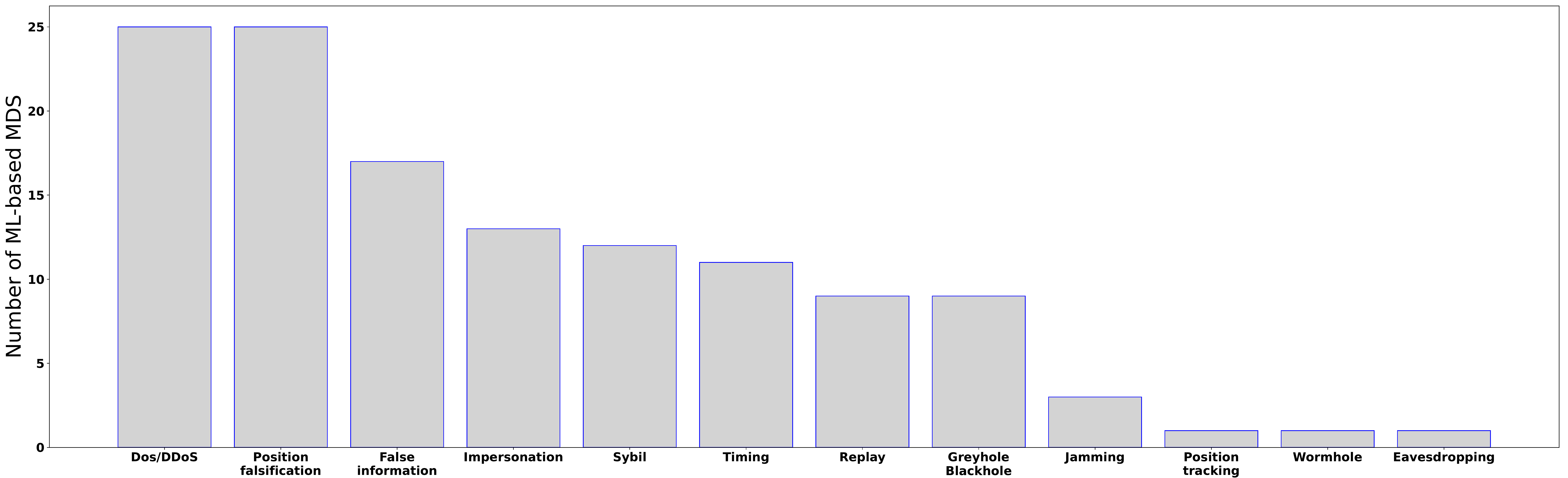}
	    \caption{Published ML-based MDSs per Attack}
		\label{fig:pubMDSperAttack}
\end{figure*}

Table~\ref{tab:MDSvsAtk} lists the publication year and the attacks targeted by the ML-based MDSs presented. As already shown in Figure~\ref{fig:pubMDSperYear}, since 2014, ML-based MDSs have witnessed an increasing interest by the research community. This is not only due to the importance of the topic and the huge advances in ML done in the few recent years., but also to the emergence of interesting datasets such as VeReMi and VeReMi extension. On the other hand, as depicted in Figure~\ref{fig:pubMDSperAttack}, the majority of proposed ML-based MDSs is targeting Dos/DDoS attacks, position falsification, false information attacks. This can be explained by their serious consequences on 5G-V2X systems.

\begin{table*}[]
\caption{ML-based MDSs vs. Attacks}
\label{tab:MDSvsAtk}
\resizebox{18cm}{12cm}{
\begin{tabular}{|l|l|c|l|l|l|l|l|l|l|l|l|l|l|}
\hline
\multicolumn{1}{|c|}{Year} & \multicolumn{1}{c|}{Work}                                                      &\rotatebox{90}{~\parbox{2cm}{Position falsification~}} & \multicolumn{1}{c|}{\rotatebox{90}{~\parbox{1.5cm}{False information~}}} & \multicolumn{1}{c|}{\rotatebox{90}{~\parbox{1.5cm}{Sybil~}}} & \multicolumn{1}{c|}{\rotatebox{90}{~\parbox{1.5cm}{Position tracking~}}} & \multicolumn{1}{c|}{\rotatebox{90}{~\parbox{1.5cm}{Dos/DDoS~}}} & \multicolumn{1}{c|}{\rotatebox{90}{~\parbox{1.5cm}{Reply~}}} & \multicolumn{1}{c|}{\rotatebox{90}{~\parbox{1.5cm}{Timing~}}} & \multicolumn{1}{c|}{\rotatebox{90}{~\parbox{1.5cm}{Greyhole /blackhole~}}} & \multicolumn{1}{c|}{\rotatebox{90}{~\parbox{1.5cm}{Jamming~}}} & \multicolumn{1}{c|}{\rotatebox{90}{~\parbox{1.5cm}{Impersonation~}}} & \multicolumn{1}{c|}{\rotatebox{90}{~\parbox{1.5cm}{Wormhole~}}} & \multicolumn{1}{c|}{\rotatebox{90}{~\parbox{1.5cm}{Eavesdropping~}}} \\ \hline
2011                       & Grover et al. (2) \cite{grover2011machine}                    & X                      &                                        &                            &                                        &                               & X                          & X                           & X                                       &                              & X                                  &                               &                                   \\ \hline
2011                       & Grover et al. (3)\cite{grover2011misbehavior}                 & X                      &                                        &                            &                                        &                               & X                          & X                           & X                                       &                              & X                                  &                               &                                   \\
\hline
\end{tabular}
}

\end{table*}

Tables~\ref{tab:SecSummaryNonIP},~\ref{tab:SecSummaryIP}, and~\ref{tab:SecSummaryboth} show the security and privacy-oriented summary for ML-based MDSs for Non-IP-based applications, for IP-based applications, and for both of them respectively. In the following, we define different criteria used in this summary.

\begin{itemize}
 \item \textbf{Type of the misbehavior}: this column mentions which type of misbehavior detect by the ML-based MDS. It can be an attack performed intentionally by a malicious node or an anomaly caused by a malfunctioning node.

\item \textbf{Unseen attacks (Yes/Maybe/No)}: this column indicates whether the ML-based can detect unseen attacks or not.  
    
    \item \textbf{Learning model}: This column indicates whether the model is built by a single node (Single) or multiple nodes (Collaborative).  

    \item \textbf{Learning mode}: depending on the learning model, this column indicates the learning mode. If the model is built by a single node, then it can be on a single server or data center. But, if the model is built collaboratively by multiple nodes, then it can be done in federated or peer-to-peer learning. 
    
    \item \textbf{Privacy preservation (Yes/No)}: this column indicates if the privacy preservation is ensured by the ML-based MDS or not.
    \item \textbf{Context-aware (Yes/No)}: this column indicates if the ML-based MDS takes into account the context parameters to change the security parameters.
    
    \item \textbf{Secure (Yes/No)}: the column indicates if the ML-based MDS is secured or not. 
    
    \item \textbf{Communication Overhead (Large/Low)}: this column indicates the communication overhead of the ML-based MDS, which is concluded according to the learning mode. According to \cite{du2020federated}, in a centralized architecture all data should be collected. and thus communication overhead is large. Similarly, in a distributed data center architecture, row data exchanges are conducted among servers, and thus the communication overhead is large. In Federated learning. Communication overhead is smaller as compared with other learning approaches. Communications are only required between the central server and each client. In peer-to-peer communication, overhead is larger than FL because more signaling overheads are needed to achieve synchronization among multiple clients. 
    
    \item \textbf{Validation (dataset/simulation)}: this column indicates how ML model is validated using dataset or through simulations.
\end{itemize}

From the analyzes of Tables~\ref{tab:SecSummaryNonIP},~\ref{tab:SecSummaryIP}, and~\ref{tab:SecSummaryboth}, we can see that almost all the proposed ML-based MDSs focus on detecting attacks instead of anomalies. In addition, since most of the proposed ML-based MDSs are supervised, the majority of MDS focuses on the detection of specific attacks instead of previously unseen attacks. However, unsupervised-based MDSs try to detect attacks by detecting the deviations from the normal behavoir. Thus, we cannot be sure that these ML-based MDSs can detect unseen attacks since the authors only evaluate them on well-known attacks and don't identify new attacks. For this reason, we put "maybe" in the "Unseen attacks" column. Moreover, in most of the works, the ML model is built based on a single node. Only a few are based on collaborative learning, and most of them leverage FL. This particular set of MDSs is privacy-preserving, unlike the others. Context-awareness is also not considered in most of the works; only two include this property. Finally, we note that all the existing works are validated using only datasets; only one uses simulations along with datasets to validate the ML model.

\begin{table*}[]
\caption{Security and privacy-oriented summary for ML-based MDSs (Non-IP-based applications)}
\label{tab:SecSummaryNonIP}

\resizebox{18cm}{!}{
\begin{tabular}{|l|l|l|l|l|l|l|l|l|l|l|}
\hline
    Work  & \begin{tabular}[c]{@{}l@{}}Misbehavior\\ Type\end{tabular} & \begin{tabular}[c]{@{}l@{}}Unseen \\ attacks\end{tabular} & Learning model & Learning mode                                                      & \begin{tabular}[c]{@{}l@{}}Privacy \\ preservation\end{tabular} & \begin{tabular}[c]{@{}l@{}}Context \\ aware\end{tabular} & Secure & \begin{tabular}[c]{@{}l@{}}Communication\\ Overhead\end{tabular} & Validation                                                      \\ \hline
So et al. \cite{so2018integrating} & Attack                                                     & No                                                        & Single         & Centralized                                                        & No                                                              & No                                                       & No     & Large                                                            & Dataset                                                         \\ \hline
Le et al.   \cite{le2019}    & Attack                                                     & No                                                        & Single         & Centralized                                                        & No                                                              & No                                                       & No     & Large                                                            & Dataset                                                         \\ \hline
Singh et al. (1) \cite{singh2019machine}  & Attack                                                     & No                                                        & Single         & Centralized                                                        & No                                                              & No                                                       & No     & Large                                                            & Dataset                                                         \\ \hline
Sharma et al. (1)  \cite{sharma2020machine}    & Attack                                                     & No                                                        & Single         & Centralized                                                        & No                                                              & No                                                       & No     & Large                                                            & Simulation                                                      \\ \hline
Montenegro et al. \cite{montenegro2020detection}  & Attack                                                     & No                                                        & Single         & Centralized                                                        & No                                                              & No                                                       & No     & Large                                                            & Simulation                                                      \\ \hline
Ecran et al. (1)  \cite{ercan2021new}      & Attack                                                     & No                                                        & Single         & Centralized                                                        & No                                                              & No                                                       & No     & Large                                                            & Dataset                                                         \\ \hline
Ecran et al. (2)   \cite{ercan2021misbehavior}  & Attack                                                     & No                                                        & Single         & Centralized                                                        & No                                                              & No                                                       & No     & Large                                                            & Dataset                                                         \\ \hline
Hawlader et al.   \cite{hawlader2021intelligent} & Attack                                                     & No                                                        & Single         & Centralized                                                        & No                                                              & No                                                       & No     & Large                                                            &  \begin{tabular}[c]{@{}l@{}}Data set\&\\ Simulation\end{tabular}                                                       \\ \hline
Okamura et al.   \cite{okamura2021misbehavior}  & Attack                                                     & Maybe                                                   & Single         & Centralized                                                        & No                                                              & No                                                       & No     & Large                                                            & Dataset                                                         \\ \hline
Grover et al. (1) \cite{9395714}  & Attack                                                     & Maybe                                                   & Single         & Centralized                                                        & No                                                              & No                                                       & No     & Large                                                            & Dataset                                                         \\ \hline
Sedar et al.   \cite{reinforcement}    & Attack                                                     & No                                                        & Single         & Centralized                                                        & No                                                              & No                                                       & No     & Large                                                            & Dataset                                                         \\ \hline
Uprety et al. \cite{uprety2021privacy}   & Attack                                                     & No                                                        & Collaborative  & Federated                                                          & Yes                                                             & No                                                       & No     & Small                                                            & Dataset                                                         \\ \hline
Sharma et al. (2) \cite{sharma2021machine, sharma2021machine2}              & Attack                                                     & No                                                        & Single  & Centralized                                                          & No                                                             & No                                                       & No     & Large                                                            & Dataset                                                         \\ \hline
Ghaleb et al. \cite{ghaleb2017effective}   & Attack                                                     & No                                                        & Single         & Centralized                                                        & No                                                              & No                                                       & No     & Large                                                            & Dataset                                                         \\ \hline
Monteuuis et al. \cite{monteuuis2018my}  & Attack                                                     & No                                                        & Single         & Centralized                                                        & No                                                              & No                                                       & No     & Large                                                            & Dataset                                                         \\ \hline
Singh et al. (2) \cite{singh2018misbehavior}   & Attack                                                     & No                                                        & Single         & Centralized                                                        & No                                                              & No                                                       & No     & Large                                                            & -                                                               \\ \hline
Gyawali et al.   \cite{gyawali2019, gyawali2020}  & Attack                                                     & No                                                        & Single         & Centralized                                                        & No                                                              & No                                                       & No     & Large                                                            & Dataset                                                         \\ \hline
Negi et al. \cite{negi2020}   & Anomaly                                                    & Maybe                                                & Single         & \begin{tabular}[c]{@{}l@{}}Data center \\ distributed\end{tabular} & No                                                              & No                                                       & No     & Large                                                            & Dataset                                                         \\ \hline

Almalki et al. \cite{almalki2021deep} & Attack                                                     & No                                                        & Single         & Centralized                                                        & No                                                              & Yes                                                      & No     & Large                                                            & Dataset                                                         \\ \hline

Gu et al. (1) \cite{gu2017support}  & Attack                                                     & No                                                        & Single         & Centralized                                                        & No                                                              & No                                                       & No     & Large                                                            & Dataset                                                         \\ \hline
Gu et al. (2) \cite{gu2017k}  & Attack                                                     & No                                                        & Single         & Centralized                                                        & No                                                              & No                                                       & No     & Large                                                            & Dataset                                                         \\ \hline

Kamel et al. (1) \cite{kamel2019misbehavior}  & Attack                                                     & No                                                        & Single         & Centralized                                                        & No                                                              & No                                                       & No     & Large                                                            & Dataset                                                         \\ \hline

Quevedo et al.   \cite{quevedo2020}  & Attack                                                     & No                                                        & Single         & Centralized                                                        & No                                                              & No                                                       & No     & Large                                                            & Dataset                                                         \\ \hline

Boualouache et al.  \cite{boualouache2021federated}  & Attack                                                     & No                                                        & Collaborative  & Federated                                                          & Yes                                                             & No                                                       & No     & Small                                                            & \begin{tabular}[c]{@{}l@{}}Data set\&\\ Simulation\end{tabular} \\ \hline
Grover et al. (2) \cite{grover2011machine}      & Attack                                                     & No                                                        & Single         & Centralized                                                        & No                                                              & No                                                       & No     & Large                                                            & Dataset                                                         \\ \hline
Grover et al. (3) \cite{grover2011misbehavior}    & Attack                                                     & No                                                        & Single         & Centralized                                                        & No                                                              & No                                                       & No     & Large                                                            & Dataset                                                         \\ \hline
Li et   al. (1) \cite{li2015svm}    & Attack                                                     & No                                                        & Single         & Centralized                                                        & No                                                              & Yes                                                      & No     & Large                                                            & Dataset                                                         \\ \hline
Zhang et al. (1)  \cite{zhang2018misbehavior}     & Attack                                                     & No                                                        & Single         & Centralized                                                        & No                                                              & No                                                       & No     & Large                                                            & Dataset                                                         \\ \hline
Eziama et al.   \cite{eziama2018malicious}  & Attack                                                     & Maybe                                                 & Single         & Centralized                                                        & No                                                              & No                                                       & No     & Large                                                            & -                                                               \\ \hline
Mahmoudi et al.   \cite{mahmoudi2019towards}      & Attack                                                     & No                                                        & Single         & Centralized                                                        & No                                                              & No                                                       & No     & Large                                                            & Dataset                                                         \\ \hline
Kamel et al. (2)  \cite{kamel2019misbehavior2}        & Attack                                                     & No                                                        & Single         & Centralized                                                        & No                                                              & No                                                       & No     & Large                                                            & Dataset                                                         \\ \hline
Alladi et al. (1)   \cite{alladi2021securing}     & Attack                                                     & No                                                        & Single         & Centralized                                                        & No                                                              & No                                                       & No     & Large                                                            & Dataset                                                         \\ \hline
Alladi et al. (2)   \cite{alladi2021artificial}  & Attack                                                     & No                                                        & Single         & Centralized                                                        & No                                                              & No                                                       & No     & Large                                                            & Dataset                                                         \\ \hline

Alladi et al. (3) \cite{alladi2021deep}   & Attack                                                     & Maybe                                              & Single         & Centralized                                                        & No                                                              & No                                                       & No     & Large                                                            & Dataset                                                         \\ \hline

Alladi et al. (4)   \cite{alladi2021deepadv} & Attack                                                     & Maybe                                               & Single         & Centralized                                                        & No                                                              & No                                                       & No     & Large                                                            & Dataset                                                         \\ \hline

Kushardianto et al. ~\cite{kushardianto20212}  & Attack                                                     & Maybe                                               & Single         & Centralized                                                        & No                                                              & No                                                       & No     & Large                                                            & Dataset                                                         \\ \hline
Gonçalves et al. \cite{gonccalves2021intelligent}   & Attack                                                     & Maybe                                               & Single         & Centralized                                                        & No                                                              & No                                                       & No     & Large                                                            & Dataset                                                         \\ \hline

\end{tabular}
}
\end{table*}

\begin{table*}[]
\caption{Security and privacy-oriented summary for ML-based MDSs (IP-based applications)}
\label{tab:SecSummaryIP}
\resizebox{18cm}{!}{
\begin{tabular}{|l|l|l|l|l|l|l|l|l|l|l|}
\hline
                                                                             Work &  \begin{tabular}[c]{@{}l@{}}Misbehavoir\\ Type\end{tabular} & \begin{tabular}[c]{@{}l@{}}Unseen \\ attacks\end{tabular} & Learning mode & Learning model                                                      & \begin{tabular}[c]{@{}l@{}}Privacy \\ preservation\end{tabular} & \begin{tabular}[c]{@{}l@{}}Context \\ aware\end{tabular} & Secure & \begin{tabular}[c]{@{}l@{}}Communication\\ Overhead\end{tabular} & Validation \\ \hline
Tan et al.   \cite{tan2018}   & Attack                                                     & Maybe                                               & Single        & Centralized                                                         & No                                                              & No                                                       & No     & Large                                                            & Dataset    \\ \hline
Singh et al. (3) \cite{singh2018}       & Attack                                                     & No                                                        & Single        & Centralized                                                         & No                                                              & No                                                       & No     & Large                                                            & Dataset    \\ \hline
Yu et al.   \cite{yu2018}   & Attack                                                     & No                                                        & Single        & Centralized                                                         & No                                                              & No                                                       & No     & Large                                                            & Dataset    \\ \hline
Sharshembiev et all~\cite{sharshembiev2021protocol}   \cite{yu2018}                                  & Attack                                                     & Maybe                                                        & Single        & Centralized                                                         & No                                                              & No                                                       & No     & Large                                                            & Dataset    \\ \hline

Gruebler et al.   \cite{gruebler2015intrusion}  & Attack                                                     & No                                                        & Single        & Centralized                                                         & No                                                              & No                                                       & No     & Large                                                            & Dataset    \\ \hline
Alheeti et al. (1)  \cite{alheeti2015}  & Attack                                                     & No                                                        & Single        & Centralized                                                         & No                                                              & No                                                       & No     & Large                                                            & Dataset    \\ \hline
Zeng et al. \cite{zeng2018senior2local}      & Attack                                                     & No                                                        & Single        & Centralized                                                         & No                                                              & No                                                       & No     & Large                                                            & Dataset    \\ \hline
Siddiqui et al.   \cite{siddiqui2019machine}     & Attack                                                     & No                                                        & Single        & Centralized                                                         & No                                                              & No                                                       & No     & Large                                                            & Dataset    \\ \hline
Acharya and Oluoch \cite{acharya2021dual}    & Attack                                                     & No                                                        & Single        & Centralized                                                         & No                                                              & No                                                       & No     & Large                                                            & Dataset    \\ \hline
Singh et al. (4) \cite{singh2019machine2}   & Attack                                                     & No                                                        & Single        & Centralized                                                         & No                                                              & No                                                       & No     & Large                                                            & Dataset    \\ \hline

Alheet et al. (2)   \cite{alheeti2016hybrid,ali2018intelligent}   & Attack                                                     & No                                                        & Single        & Centralized                                                         & No                                                              & No                                                       & No     & Large                                                            & Dataset    \\ \hline
Kim et al. \cite{kim2017}         & Attack                                                     & No                                                        & Single        & Centralized                                                         & No                                                              & No                                                       & No     & Large                                                            & Dataset    \\ \hline
Zhang et al. (2) \cite{zhang2018distributed}   & Attack                                                     & No                                                        & Collaborative & \begin{tabular}[c]{@{}l@{}}Peer-to-peer\\ distributed\end{tabular}  & (Yes)                                                           & No                                                       & No     & Large                                                            & Dataset    \\ \hline
Ghaleb et al. \cite{a2020misbehavior}    & Attack                                                     & No                                                        & Collaborative & \begin{tabular}[c]{@{}l@{}}Peer-to-peer \\ distributed\end{tabular} & No                                                              & No                                                       & No     & Large                                                            & Dataset    \\ \hline
Ashraf et al.   \cite{ashraf2020novel}        & Attack                                                     & Maybe                                              & Single        & Centralized                                                         & No                                                              & No                                                       & No     & Large                                                            & Dataset    \\ \hline
Shu et al.   \cite{shu2020collaborative}    & Attack                                                     & No                                                        & Collaborative & Centralized                                                         & No                                                              & No                                                       & No     & Large                                                            & Dataset    \\ \hline
Li et al. (2)   \cite{li2020transfer}   & Attack                                                     & No                                                        & Single        & Centralized                                                         & No                                                              & No                                                       & No     & Large                                                            & Dataset    \\ \hline
Bangui et al.   \cite{bangui2021hybrid}    & Attack                                             & Yes                                                       & Single        & Centralized                                                         & No                                                              & No                                                       & No     & Large                                                            & Dataset    \\ \hline
Yang et al.   \cite{yang2021mth}        & Attack                                            & Yes                                                       & Single        & Centralized                                                         & No                                                              & No                                                       & No     & Large                                                            & Dataset    \\ \hline
Khan et al. \cite{khan2021enhanced}  & Attack                                             & Yes                                                       & Single        & Centralized                                                         & No                                                              & No                                                       & No     & Large                                                            & Dataset    \\ \hline
Liu et al.   \cite{liu2021blockchain}   & Attack                                                     & No                                                        & Collaborative & Federated                                                           & Yes                                                             & No                                                       & Yes    & Small                                                            & Dataset    \\ \hline
Rahal et al. \cite{rahal2022antibotv}   & Attack                                                     & No                                                        & Single & Centralized                                                          & No                                                             & No                                                        & No    & Large                                                            & Dataset    \\ \hline
\end{tabular}
}
\end{table*}

\begin{table*}[]
\caption{Security and privacy-oriented summary for ML-based MDSs (both)}
\label{tab:SecSummaryboth}
\resizebox{18cm}{!}{
\begin{tabular}{|l|l|l|l|l|l|l|l|l|l|l|}
\hline
                                                                             Work & \begin{tabular}[c]{@{}l@{}}Misbehavoir\\ Type\end{tabular} & \begin{tabular}[c]{@{}l@{}}Unseen \\ attacks\end{tabular} & Learning mode & Learning model                                                      & \begin{tabular}[c]{@{}l@{}}Privacy \\ preservation\end{tabular} & \begin{tabular}[c]{@{}l@{}}Context \\ aware\end{tabular} & Secure & \begin{tabular}[c]{@{}l@{}}Communication\\ Overhead\end{tabular} & Validation \\ \hline

Karagiannis et al.   \cite{karagiannis2018}   & Attack                                            & Yes                                                       & Single        & Centralized                                                         & No                                                              & No                                                       & No     & Large                                                            & Dataset    \\ \hline
Lyamin et al.   \cite{lyamin2018ai}   & Attack                                             & Yes                                                       & Single        & Centralized                                                         & No                                                              & No                                                       & No     & Large                                                            & Dataset    \\ \hline
Abhishek et Gurusamy \cite{abhishek2021jade}   & Attack                                             & Yes                                                       & Single        & Centralized                                                         & No                                                              & No                                                       & No     & Large                                                            & Dataset    \\ \hline

\end{tabular}
}
\end{table*}

\subsection{ML-Oriented summary}
Tables~\ref{tab:MLSummaryNonIP},~\ref{tab:MLSummaryIP}, and~\ref{tab:MLSummaryboth} show the ML-oriented summary for MDS for Non-IP-based applications, for IP-based applications, and for both of them respectively. In the following, we define different criteria used in this summary.

\begin{itemize}

\item \textbf{ML method}: this column mentions the ML method used to train the ML model including supervised, unsupervised, hybrid (combines supervised and unsupervised methods), and reinforcement learning.
    
\item \textbf{Dataset}: This column indicates which dataset is used to train the ML model. The name of the dataset is mentioned if it is publicly available. Otherwise, the network simulator used to generate the dataset is mentioned. 

\item \textbf{ML Task}: according to the used ML method, this column indicates the type of ML task. For supervised learning: the task can be regression, binary classification, or multi-class classification. For unsupervised learning: the task can be anomaly detection or clustering. For reinforcement learning, the task can be Markov Decision Process or Q-learning. 

\item \textbf{Update model (Yes/No)}: this column indicates whether the ML model is updated overtime or not.   

\item \textbf{ML Algorithm}: this column mentions the ML algorithms used to train the model and whether these algorithms are traditional or based on deep learning.

\item \textbf{Metrics}: this column indicates which are the metrics used to evaluate the ML model.

\item \textbf{Inference loc}: this column mentions the location where the ML model is deployed after the validation tests. This could be vehicles, edge nodes, RSUs, or the Cloud.
\end{itemize}

From the analyzes of Tables~\ref{tab:MLSummaryNonIP},~\ref{tab:MLSummaryIP}, and~\ref{tab:MLSummaryboth} , we see that most of ML-based MDSs use a supervised ML method. Also, only an ML-based MDS uses a hybrid ML method combining unsupervised and supervised methods. Thus, most ML tasks are classification tasks that use either binary classifiers or binary classifiers.  We also note unsupervised tasks are mostly anomaly detection tasks.

From the perspective of the used dataset, we can conclude that ML-based MDS for non-IP-based based applications (Table~\ref{tab:MLSummaryNonIP}) are generated using network simulators. In addition, the majority of authors prefer to use public datasets (VeReMi and VeReMi extended) than generating their own datasets. But, ML-based MDSs for IP-based applications have mostly used public datasets generated using computer networks testbeds.

On other hand,   most of the works built their ML models using traditional ML algorithms. However, the latest works have started to focus on more deep learning algorithms. We also see that the used evolution metrics are different from an ML-based MDS to another and mainly depend on the authors' perspectives, which are dependent on what to demonstrate).  In addition, in most ML-based MDS the models are not updated after their deployment. Only a few works explicitly mention the update process, especially where the update is done by design like in FL-based MDS. Finally, the majority of work doesn't mention the inference location, but for the whom who mention is mostly vehicles.


\begin{table*}[]
\caption{ML-oriented summary for ML-based MDSs (Non-IP-based applications}
\label{tab:MLSummaryNonIP}

\resizebox{18cm}{!}{

}
\end{table*}

\begin{table*}[]
\caption{ML-oriented summary for ML-based MDSs (IP-based applications)}
\label{tab:MLSummaryIP}
\resizebox{18cm}{!}{
%
}
\end{table*}

\begin{table*}[]
\caption{ML-oriented summary for ML-based MDSs (both)}
\label{tab:MLSummaryboth}
\resizebox{18cm}{!}{
%
                                                                                                                       & Detection probability                                                                                             & Vehicle                                                            \\ \hline

\end{tabular}
}
\end{table*}

\section{Lessons learned and recommendations}
\label{sec:lessons}
Interesting lessons and some recommendations could be concluded from the results and analyses presented in the previous section. First of all, the proposed ML-based MDSs are strongly dependent on the training data set. Specifically, the attacks detected by these MDS only cover the attacks included in datasets unlike what is described in the title and the abstracts of most of the research papers. These latter give the impression that are proposing solutions that address all the attacks instead of specific attacks. We strongly recommend authors to be concise when presenting their ML-based MDS. We also recommend propose holistic security frameworks that can integrate different ML-based MDS for covering various existing attacks including unseen attacks.
Moreover, we notice that the proposed ML-based MDSs are generally incomparable due to the absence of benchmark datasets and the unuse of unified evaluation metrics. Indeed, an important part of research papers generates their datasets, which makes it difficult to reproduce their results and compare them with other solutions. The rest of the research papers use public data sets such as VeReMi and VeRemi extension. However, these datasets are difficult to be appointed dataset as benchmark datasets due to several data described in the open issues section (\ref{subsec:dataset}). To this end, we recommend gathering efforts for defining unified benchmark datasets. We also recommend specifying a common evaluation framework consisting of all metrics used to evaluate and compare ML-based MDS. We encourage researchers to reproduce the results of different ML-based MDS and compare the results. This task cannot be easy since the majority of authors didn't well explain the methodoly and the parameters of the ML models. Thus, we recommend research working in this field to increase their knowledge in ML, include the required parameters to reproduce their results, and preferability make the implementation publically available. Furthermore, most of the proposed ML-based MDSs ignore the deployment phase of the ML-based MDS.  Indeed, in these works,  the inference location is not even mentioned and the update of the ML model after deployment is not considered. Thus, we recommend paying careful attention to the deployment phase, which has an important impact not only on the detection rate but also on the feasibility of the ML-based MDS. New indirect evaluation metrics should be included for the study of the ML-based MDSs such as the size of the ML model and the inference time (time to detect the attack). The inference location should be carefully studied not only from the detection perspective but also from security and privacy perspectives. The mechanisms to update the models should be defined to prevent loss of accuracy with time. In this direction, collaborative MDS offers interesting opportunities to smoothly update the ML model and provide privacy preservation. For this reason, we recommend promoting research in this direction by exploring more advanced ML concepts such as online learning and reinforcement learning. However, ML-based MDS including collaborative ones, are still facing a range of security threats such as adversarial attacks and poisoning. Our analyzes identify only one work that considers the security of the ML-based MDS. To this end, we believe that the security of ML-based MDS is an urgent issue that requires concerted efforts.


\section{Open research issues}
\label{sec:issues}

\begin{figure}[!ht]
		\centering
		\includegraphics[width=9cm,height=3.5cm]{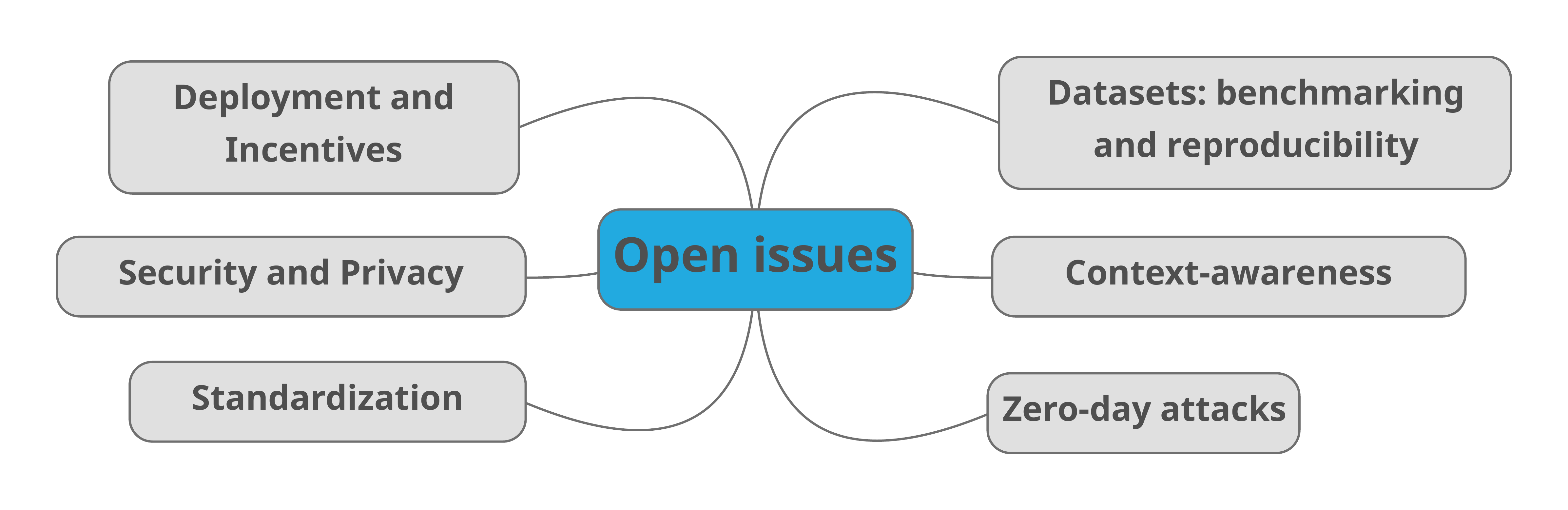}
	    \caption{Open issues for ML-based MDSs}
		\label{fig:openissues}
\end{figure}

Several parameters involve in building effective ML-based MDSs such as the quality of datasets and the used ML algorithms. However, although considerable efforts have been made, several open issues are still needing more attention to achieve the aimed ML-based MDS. Some of these issues, illustrated in Figure~\ref{fig:openissues}, are discussed in this section.

\subsection{Datasets: benchmarking and reproducibility} \label{subsec:dataset}

As already discussed in the learned lessons section, the absence of benchmark datasets is a major open issue to reproduce results of existing ML-based MDSs and compare them. However, we believe that it is difficult to appoint benchmark datasets for three following main reasons: (i) the absence of standardized procedures that clearly define V2X attacks scenarios; (ii) a part of the used datasets are generated using vehicular network simulators, which cannot capture of all the realistic parameters; (iii) the rest of datasets were generated using computer network testbeds, which doesn't consider the mobility of vehicles; an important feature of 5G-V2X. To this end, we believe that efforts should be gathered for defining clear standardized V2X attack scenarios to generate unified benchmark datasets using realistic vehicular testbeds. A common evaluation framework consisting of all metrics used to evaluate and compare ML-based MDS should also be specified. Besides, the reproducibility of results is another issue since authors tend to neither well describe ML parameters nor make their source codes public. For this reason, we encourage authors to adopt a result reproducibility methodology. We also encourage researchers to reproduce the results of different ML-based MDSs and compare the results.  


\subsection{Zero-day attacks}
Zero-days attacks are vectors of unseen attacks that appear over time due to the evolution of technologies (i.e. network slicing \cite{mun2021secure}) and attacker strategies \cite{cho2020toward}.  Detecting zero-days is still an open issue. On one hand, the majority of existing ML-based MDSs were built on a supervised approach, which allows only detecting specific attacks. On the other hand, the rest of ML-based MDSs use unsupervised models, which are built based on normal data only. Thus, they can detect anomalies but cannot identify them. To this end, detecting zero-days is very challenging.  New emerging hybrid semi-automatic frameworks including humans in the detection loop are promising \cite{sedjelmaci2021trusted}.  However, the ultimate goal is to detect zero-day attacks automatically under the umbrella of the zero-touch paradigm \cite{benzaid2020ai}.

\subsection{Context-awareness}

Almost all the existing ML-based MDSs leverage the direct parameters to detect attacks ignoring indirect parameters that have also an influence on the attack detection accuracy. For example, detecting message greyhole/blackhole attacks needs monitoring of message exchanges between nodes vehicles. However, the messages could be suppressed intentionally due to environmental characteristics such as obstacles and interference. Thus, to detect such behavior other indirect parameters such as channels status,  temperature, and speed should also be considered. Although two ML-based MDSs \cite{li2015svm,almalki2021deep}  consider the context, their contributions are still limited since indirect parameters are not included in the ML training. Besides, SDN approaches to change security parameters according to the context are interesting \cite{shu2020collaborative,boualouache2020sdn}.

\subsection{ Security and Privacy}

As discussed in the analyzes section, the majority of existing ML-based MDSs are centralized where the datasets are collected and the ML model is trained in one location. This exposes them to serious security and privacy issues for the following reasons: (i) these solutions are suffering from the single-point-of-failure attacks since the model is trained on a single location; and (ii) collecting datasets by one entity could have several privacy violations since datasets may contain sensitive information about the behavior and movement patterns of V2X nodes. Collaborative FL-based MDSs~\cite{uprety2021privacy,boualouache2021federated, liu2021blockchain} came to partially address privacy issues since datasets in these systems are shared among learning nodes instead of centrally stored. However, security issues are multiplying with the number of components in collaborative systems. In addition, FL-based MDSs are still suffering from single-point-of-failure attacks since the global model is aggregated and calculated in one FL server.  To this end, we believe that blockchain could be an efficient technology to secure ML-based MDSs \cite{liu2021blockchain}. However, there are still some problems that need to be addressed in the blockchain design to achieve this aim such as the consensus algorithms and optimizing smart contracts. 

Besides, like other systems based on ML, ML-based MDS are also suffering from the adversarial ML \cite{sharma2019attacks, qayyum2020securing}. The adversarial ML is a set of techniques that try to exploit models by making use the information obtained from models to launch advanced attacks. For example, in FL  even the datasets aren't shared, learning nodes are still sending small updates to the FL server. This information can be used by an attacker to infer sensitive information about the model and thereby launch attacks to poison the model \cite{lyu2020threats}. Although recent work has addressed the adversarial ML attacks issue~\cite{talpur2021adversarial},  we believe that this issue still needs careful attention.

\subsection{Deployment and Incentives}

Building a successful ML-based MDS is not only dependent on the development and validation phases but also the deployment phase. In addition to the datasets and the accuracy model,  ML-based MDSs should be developed to fit the V2X environment (software and hardware) on which these systems will be deployed. Thus, ML-based MDSs should be developed with the end in mind .i.e metrics such as the size of the model and the processing resources required to run the models should be considered in the evaluation part. Besides,  the placement of ML-based MDS components should be studied to provide an early attack detection and a rapid reaction while protecting them from vulnerabilities.  We believe that the deployment of ML-based MDSs is an open issue that requires efforts from both research and industry.

Besides, since the deployment of  ML-based MDS in V2X nodes will consume storage and processing resources, the manager of these nodes might have an objection to deploying ML-based MDS. Thus, the incentive issue should also be addressed to ensure the continuity of ML-based MDS service. Some works have started interesting in incentive modeling using game theory frameworks~\cite{xing2019trust}. However, we believe that more efforts can still be done in this is direction.


\subsection{Standardization}
The first efforts on standardization of MDSs are ongoing \cite{ansari2021v2x}. ETSI has recently published a technical report on the pre-standardization study of V2X misbehavior detection \cite{etsi3}. Although several detection techniques are mentioned in this report, the role of ML is not well emphasized. We believe that standardization bodies should put more focus on defining a toolbox for developing and validating ML-based MDSs. Consequently, we can identify several standardization opportunities: 1) defining attack scenarios in complementary with ETSI TR 102 893 \cite{etsi4}; (2) defining benchmark datasets; (4) defining validation KPIs; (5) specifying evaluation metrics;  and (6) establish clear validation procedures.  On the other hand, standardization bodies should organize plugtests events (.i.e \cite{plugtests}) that gather several stakeholders for testing and validating ML-based MDSs with reporting relevant results as is the case \cite{etsi5,etsi6,etsi7}.



\section{Conclusion}
\label{sec:conclusion}

Misbehavior Detection Systems (MDS) are key building blocks for securing 5G-V2X networks. Machine Learning (ML) is an inevitable part of the design of these systems. An increasing effort is ongoing for providing effective ML-based MDSs. In this paper, we surveyed and classified relevant ML-based MDSs for 5G-V2X. We also analyzed and discussed them from security and ML perspectives. Finally, we gave some learned lessons and shed light on open research and standardization issues for building effective ML-based MDSs.

\section*{Acknowledgment}
This work was supported by the 5G-INSIGHT bilateral project, (ANR-20-CE25-0015-16), funded by the Luxembourg National Research Fund (FNR), and by the French National Research Agency (ANR).

\ifCLASSOPTIONcaptionsoff
  \newpage
\fi

\bibliographystyle{IEEEtran}
\bibliography{IEEEabrv,survey_refs}

\end{document}